\documentclass[reprint,
 amsmath,amssymb,
 aps,
prd,
showkeys,
]{revtex4-1}

\usepackage{graphicx}
\usepackage{dcolumn}
\usepackage{bm}
\usepackage{hyperref}
\usepackage{mathtools,cancel}
\usepackage{tikz-feynman}
\usepackage{slashed}
\usepackage{xcolor}
\usepackage{float}
\usepackage{appendix}
\usepackage{ulem}

\begin{document}

\title{Fluctuating temperature and baryon chemical potential in heavy-ion collisions and the position of the critical end point in the effective QCD phase diagram }
\author{Alejandro Ayala$^{1,2}$, Saul Hern\'andez-Ortiz$^{1,3}$, L. A. Hern\'andez$^{1,2,4}$, V\'ictor Knapp-P\'erez$^{1}$ and R. Zamora$^{5,6}$}
\affiliation{%
$^1$Instituto de Ciencias Nucleares, Universidad Nacional Aut\'onoma de M\'exico, Apartado Postal 70-543, CdMx 04510, Mexico.\\
$^2$Centre for Theoretical and Mathematical Physics, and Department of Physics, University of Cape Town, Rondebosch 7700, South Africa.\\
$^3$ Institute for Nuclear Theory, University of Washington, Seattle, WA, 98195, USA. \\
$^4$Facultad de Ciencias de la Educaci\'on, Universidad Aut\'onoma de Tlaxcala, Tlaxcala, 90000, Mexico. \\
$^5$Instituto de Ciencias B\'asicas, Universidad Diego Portales, Casilla 298-V, Santiago, Chile.\\
$^6$Centro de Investigaci\'on y Desarrollo en Ciencias Aeroespaciales (CIDCA), Fuerza A\'erea de Chile, Casilla 8020744, Santiago, Chile.
}%

\begin{abstract}
We use the linear sigma model with quarks to locate the critical end point in the effective QCD phase diagram accounting for fluctuations in temperature and quark chemical potential. For this purpose, we use the non-equilibrium formalism provided by the superstatistics framework. We compute the effective potential in the high- and low-temperature approximations up to sixth order and include the contribution of ring diagrams to account for plasma screening effects. We fix the model parameters from relations between the thermal sigma and pion masses  imposing a first order phase transition at zero temperature and a finite critical value for the baryon chemical potential that we take of order of the nucleon mass. We find that the CEP displacement due to fluctuations in temperature and/or quark chemical potential is almost negligible.
\end{abstract}

\maketitle


\section{Introduction}\label{sec1}


The study of the transition describing the phase change from nuclear to quark-gluon matter with increasing temperature ($T$) and baryon chemical potential ($\mu_B$), constitutes one of the most active fields of research of modern high-energy nuclear physics.
The description of this transition is encoded in the so called QCD phase diagram~\cite{Fukushima} represented by a temperature vs. baryon chemical potential plane, where the different phases of strongly interacting  matter can be identified. From the theoretical side, efforts to describe this diagram have been carried out employing a wide range of tools such as finite energy sum rules, Dyson-Schwinger equations, functional renormalization, holography and effective models~\cite{values, Ayala-Dominguez, Xin, Fu:2019hdw, Fischer, Fischer2, Herbst, Benic, Lu, Shi, Contrera, Cui, Datta, Knaute, Antoniou, Rougemont, Fang:2018axm, RMF, zamora1, zamora2,zamora3,Schaefer,Marczenko}. Lattice QCD (LQCD) calculations are also a useful technique~\cite{taylor0,taylor1,taylor2,taylor3,deForcrand:2008,Bonati:2018,Bazavov2018,Borsanyi,Guenther2018,Kashiwa,Cea,Bonati,Bonati:2015,Bellwied,Cea:2015,Bazavov2}, although these cannot be used to access large values of $\mu_B$, given the severe sign problem~\cite{sign}. 

On the other hand, relativistic heavy-ion collisions provide the experimental tool to access the properties of the QCD phase diagram. In recent years the STAR BES-I program has analyzed data from nuclear collisions in the energy range $200\ \text{GeV}> \ \sqrt{s_{NN}}>7.7 \ \text{GeV}$~\cite{BES1}, to explore deeper into the QCD phase diagram reaching nuclear matter at higher densities. Also, new experiments will soon enter into operation, providing data at lower collision energies with higher luminosity to better study the properties of baryon-rich matter~\cite{futureExp}.


It is well known that  fluctuations play a relevant role for the analysis and interpretation of heavy-ion data~\cite{Wilk2,Wilk3,Wilk4,Rybczynski:2014cha,
Wong:2015mba,Wilk5,Bialas:2015pla,Bhattacharyya:2015hya,Bialas:2015oua,Rozynek:2016ykp,Tripathy:2016hlg,Grigoryan:2017gcg,Bhattacharyya:2017hdc,Khuntia:2017ite,Tripathy:2017nmo,Ishihara:2017txj,Wilk:2018kvg}. In particular, when $T$ and/or $\mu_B$ are not uniform over the entire reaction volume, fluctuations from the average values can be accounted for using the so-called superstatistics scenario~\cite{Beck} where a non-extensive behavior naturally emerges due to these fluctuations. Indeed,
at the onset, the reaction starts off from regions the size of overlapping nucleon pairs. If overall thermalization is to be achieved, it seems natural to assume that these regions form subsystems from where thermalization spreads later over the entire reaction volume. In this scenario, the temperature and chemical potential between subsystems may not be the same. Thus, a superposition of two statistics needs to be considered: one in the usual Gibbs-Boltzmann sense for particles in each subsystem and another one for the probability to find particular values for $T$ and $\mu_B$ for different subsystem.

In this work, we study the QCD phase diagram implementing superstatistics with fluctuations both in $T$ and $\mu_B$. We concentrate on the location of the critical end point (CEP) using the linear sigma model with quarks (LSMq) as an effective model to find the boundaries on the phase diagram from the chiral symmetry restoration/breaking point of view. We show that when conditions for a first order phase transition are enforced to happen for a critical value of $\mu_B=\mu_B^c$ at $T=0$, the position of the CEP varies little as compared to an analysis where one only follows the evolution of the crossover transition from high $T$ without requiring a first order phase transition at $\mu_B^c$. 

The work is organized as follows. In Sec.~\ref{sec2} we review the superstatistics ideas when temperature and chemical potential fluctuate between different subsystems. We expand the generalized Boltzmann factor to first order in $1/N$, where $N$ is the number of subsystems that make up the whole system. In Sec.~\ref{sec3} we discuss the calculation of the effective potential in the LSMq, including screening effects together with the effective couplings computed at finite $T$ and $\mu_B$. In order to explore a wide region in the phase diagram, the effective potential is computed analytically in two regimes; first  at low temperature and high chemical potential, and then at high temperature and low chemical potential. 
In Sec.~\ref{sec4} we spell out the conditions that give rise to the equations to find the values of the model coupling constants. In Sec.~\ref{sec5} we use these couplings to compute the critical $T$ and $\mu_B$ that define the transition curves and locate the CEP, for the cases of superstatistics with varying number of initial subsystems and comparing to the result in the thermodynamic limit. Finally, we summarize and conclude in Sec.~\ref{sec6}. We reserve for the appendices the explicit computation of the 
vacuum stability conditions and the temperature and baryon chemical potential dependence of the effective coupling constants.

\section{Superstatistics}\label{sec2}

To include the superstatistics effects, we closely follow Ref.~\cite{AHHLZ}. accounting also as an extra ingredient fluctuations in the ratio $\mu_B/T$. Recall that for a thermodynamic system which contains an space fluctuating intensive variable $\beta$, such as the inverse temperature or the product of the chemical potential and the inverse temperature, 
one may consider the full system as made up of subsystems where $\beta$ is constant with a local Boltzmann factor $e^{-\beta\Hat{H}}$, with $\hat{H}$ being the Hamiltonian. A generalized Boltzmann factor $B(\hat{H})$ can be defined as the normal Boltzmann factor $e^{-\beta\hat{H}}$ weighted by a distribution function $f(\beta)$,
\begin{equation}
    \label{ec:Boltzmanngeneralidado}
    B(\hat{H})=\int_{0}^{\infty}f(\beta)e^{-\beta\hat{H}}d\beta \; .
\end{equation}

Let us consider the two fluctuating intensive variables $\beta=1/T$ and $\eta=\mu/T$. The local Boltzmann factor is $e^{-\beta(\hat{H}-\mu\hat{Q})}$, where $\hat{Q}$ is the number operator and we have assumed that each subsystem is part of a Grand Canonical ensemble. The generalized Boltzmann factor may be defined in a similar fashion as in Eq.~(\ref{ec:Boltzmanngeneralidado}) with probability distribution $F$ that depends on both variables, namely, $F(\beta,\eta)$. Assuming that the two variables are statistically independent, then $F(\beta, \eta)=f(\beta)g(\eta)$ and the generalized Boltzmann factor becomes
\begin{equation}
    \label{ec:CanoncalGeneralized}
    B(\hat{H},\hat{Q})=\int_{0}^{\infty}\int_{0}^{\infty}f(\beta)g(\eta)e^{-(\beta\hat{H}-\eta\hat{Q})}d\beta d\eta \; .
\end{equation}
To carry out the calculations, we consider that $f(\beta)$ and $g(\eta)$ correspond to a $\chi^{2}$ distribution function ${\mathcal{F}}$ given by
\begin{equation}
    \label{ec:chi^2}
    {\mathcal{F}}(x)=\frac{1}{\Gamma(N/2)}\bigg(\frac{N}{2x_{0}}\bigg)^{N/2}x^{N/2-1}e^{-Nx/2x_{0}},
\end{equation}
where $\Gamma$ is the gamma function, $N$ represents the number of subsystems in the full system, $x$ corresponds either to $T$ or $\eta$. The average of the random variable $x$ is given by 
\begin{equation}
    \label{ec:averagebeta}
    x_{0}\equiv \int_{0}^{\infty}x f(x)dx=\langle x \rangle,
\end{equation}
The $\chi^{2}$ describes the distribution of the sum of $N$ random variables $X_{i}$, each of which are in turn distributed obeying a Gaussian distribution. Therefore, the $\chi^{2}$ distribution is well-suited to describe random variables with positive definite values.

To evaluate Eq.~(\ref{ec:CanoncalGeneralized}) with the distribution functions given by Eq.~(\ref{ec:chi^2}), we can use that
\begin{align}
    \label{ec:integral1}
    \int_{0}^{\infty} & \frac{d \alpha}{\Gamma(N/2)}\bigg(\frac{N}{2\alpha_{0}}\bigg)^{N/2}\alpha^{\frac{N}{2}-1}e^{-\frac{N\alpha}{2\alpha_0}}e^{-\alpha\hat{A}}\nonumber \\
    &=\bigg(1+\frac{2}{N}\alpha_{0}\hat{A}\bigg)^{-\frac{N}{2}},
\end{align}
where $\hat{A}$ is any operator. The Taylor series expansion of Eq.~(\ref{ec:integral1}) is
\begin{equation}
    \label{ec:alphaseries}
   \bigg(1+\frac{2}{N}\alpha_{0}\hat{A}\bigg)^{-\frac{N}{2}}\approx \bigg[1+\frac{1}{N}\alpha_{0}^{2}\hat{A}^{2}+...\bigg]\times e^{-\alpha_{0}\hat{A}}.
\end{equation}
Therefore, using that $\hat{Q}$ commutes with $\hat{H}$, Eq.~(\ref{ec:CanoncalGeneralized})  can be written as
\begin{align}
    \label{ec:seriescanonical}
    B(\hat{H},\hat{Q})&=\bigg[1+\frac{1}{N}\beta_{0}^{2}\hat{H}^{2}+...\bigg]\times e^{-\beta_{0}\hat{H}}\nonumber \\
    &\times \bigg[1+\frac{1}{N}\eta_{0}^{2}\hat{Q}^{2}+...\bigg]\times e^{\eta_{0}\hat{Q}}.
\end{align}
Expanding Eq.~(\ref{ec:seriescanonical}), up to first order in $1/N$, we get the generalized Boltzmann factor
\begin{eqnarray}
    B(\hat{H},\hat{Q})&=&\bigg( 1+\frac{1}{N}\beta_{0}^{2}\hat{H}^{2}+\frac{1}{N}\eta_{0}^{2}\hat{Q}^{2}\bigg) e^{-(\beta_{0}\hat{H}-\eta_{0}\hat{Q})}\nonumber\\
   &=&\bigg( 1 +\frac{\beta_{0}^{2}}{N} \bigg(\frac{\partial}{\partial\beta_{0}}\bigg)^{2}+\frac{\eta_{0}^{2}}{N} \bigg(\frac{\partial}{\partial\eta_{0}}\bigg)^{2}\bigg) e^{-(\beta_{0}\hat{H}-\eta_{0}\hat{Q})}.\nonumber\\
   \label{ec:derivatives}
\end{eqnarray}
Recall that the partition function is $Z=Tr[B(\hat{H},\hat{Q})]$, therefore, we explicitly get
\begin{equation}
\label{ec:partitionfunction1}
    Z=\bigg( 1 +\frac{\beta_{0}^{2}}{N} \bigg(\frac{\partial}{\partial\beta_{0}}\bigg)^{2}+\frac{\eta_{0}^{2}}{N} \bigg(\frac{\partial}{\partial\eta_{0}}\bigg)^{2}\bigg) Z_{0},
\end{equation}
with 
\begin{equation}
\label{ec:partitionfunction0}
    Z_{0}=e^{-\beta_{0} \Omega V^{eff}(\beta_{0},\eta_{0})},
\end{equation}
where $\Omega$ and $V^{eff}(\beta_{0},\eta_{0})$ represent the volume and the effective potential, respectively.

Changing the variable $\beta_{0}$ to $T_{0}$, the partition function is explicitly obtained in terms of $T_{0}$ and $\eta_{0}$,
\begin{align}
    \label{ec:boxedz0}
    Z&=\bigg[1+\frac{1}{NT_{0}^{2}}\bigg(2T_{0}^{3}\frac{\partial}{\partial T_{0}}+T_{0}^{4}\frac{\partial^{2}}{\partial T_{0}^{2}}\bigg)+\frac{\eta_{0}^{2}}{N}\frac{\partial^{2}}{\partial\eta_{0}^{2}}\bigg]Z_{0}\nonumber \\
    &=Z_{0}\bigg[ 1 + \frac{2T_{0}}{NZ_{0}}\bigg(\frac{\partial Z_{0}}{\partial T_{0}} + \frac{T_{0}}{2}\frac{\partial^{2}Z_{0}}{\partial T_{0}^{2}}\bigg) + \frac{\eta_{0}^{2}}{NZ_{0}}\frac{\partial^{2}Z_{0}}{\partial\eta_{0}^{2}}\bigg].
\end{align}

To obtain the effective potential with superstatistics corrections, recall that 
\begin{equation}
    \label{ec:vefffsup}
    Z=e^{-\frac{\Omega}{T_{0}}V^{eff}_{sup}(T_{0},\eta_{0})}.
\end{equation}
Thus
\begin{equation}
    \label{ec:veffsup2}
 V^{eff}_{sup}(T_{0},\eta_{0})=-\frac{T_{0}}{\Omega}\ln[Z].
\end{equation}
The logarithm of the partition function in  Eq.~(\ref{ec:boxedz0}) is given by
\begin{align}
\label{ec:lnz}
    \ln[Z]&=\ln[Z_{0}]\nonumber \\
    &+\ln \bigg[ 1 + \frac{2T_{0}}{NZ_{0}}\bigg(\frac{\partial Z_{0}}{\partial T_{0}} + \frac{T_{0}}{2}\frac{\partial^{2}Z_{0}}{\partial T_{0}^{2}}\bigg) \nonumber \\
    &+ \frac{\eta_{0}^{2}}{NZ_{0}}\frac{\partial^{2}Z_{0}}{\partial\eta_{0}^{2}}\bigg].
\end{align}
Substituting Eq.~(\ref{ec:lnz}) into Eq.~(\ref{ec:veffsup2}), we obtain the effective potential with corrections due to fluctuations in $\beta$ and $\eta$
\begin{align}
  V^{eff}_{sup}(T_{0},\eta_{0})
  &=-\frac{T_{0}}{\Omega}\ln[Z_{0}]\nonumber\\
  &-\frac{T_{0}}{\Omega}\ln\bigg[ 1 + \frac{2T_{0}}{NZ_{0}}\bigg(\frac{\partial Z_{0}}{\partial T_{0}} + \frac{T_{0}}{2}\frac{\partial^{2}Z_{0}}{\partial T_{0}^{2}}\bigg) \nonumber \\
  &+ \frac{\eta_{0}^{2}}{NZ_{0}}\frac{\partial^{2}Z_{0}}{\partial\eta_{0}^{2}}\bigg] 
  \label{Veffsuper}
\end{align}
Finally, using Eq.~(\ref{ec:partitionfunction0}) into Eq.~(\ref{Veffsuper}), we obtain the expression for the effective potential in terms of $T_{0}$ and $\eta_{0}$
\begin{align}
    \label{ec:vefffinal}
    V^{eff}_{sup}(T_{0},\eta_{0})&=V^{eff}(T_{0},\eta_{0})\nonumber \\
    &-\frac{T_{0}}{\Omega}\ln \bigg[ 1 + \frac{2T_{0}}{NZ_{0}}\bigg(\frac{\partial Z_{0}}{\partial T_{0}}+ \frac{T_{0}}{2}\frac{\partial^{2}Z_{0}}{\partial T_{0}^{2}}\bigg)\nonumber \\
    &+ \frac{\eta_{0}^{2}}{NZ_{0}}\frac{\partial^{2}Z_{0}}{\partial\eta_{0}^{2}}\bigg]. 
\end{align}
We now proceed to compute the effective potential $V^{eff}$ using the LSMq.

\section{Effective potential}\label{sec3}

In order to explore the effective QCD phase diagram from the chiral symmetry prospective, we work with an effective model that takes into account spontaneous symmetry breaking/restoration at finite temperature and density; the LSMq. The dynamical degrees of freedom consist of the lightest quarks together with the pions and the sigma. The Lagrangian is given by
\begin{align}
   \mathcal{L}&=\frac{1}{2}(\partial_\mu \sigma)^2  + \frac{1}{2}(\partial_\mu \vec{\pi})^2 + \frac{a^2}{2} (\sigma^2 + \vec{\pi}^2) - \frac{\lambda}{4} (\sigma^2 + \vec{\pi}^2)^2 \nonumber \\
   &+ i \bar{\psi} \gamma^\mu \partial_\mu \psi -g\bar{\psi} (\sigma + i \gamma_5 \vec{\tau} \cdot \vec{\pi} )\psi,
\label{lagrangian}
\end{align}
where $\psi$ is an SU(2) isospin doublet of $u$ and $d$-quarks, $\vec{\pi}=(\pi_1, \pi_2, \pi_3 )$ is an isospin triplet of pions and $\sigma$ is an isospin singlet. $\lambda$ is the boson's self-coupling, $g$ is the fermion-boson coupling and $a^2>0$ is the mass parameter. To allow for an spontaneous breaking of symmetry, we let the $\sigma$ field to develop a vacuum expectation value $v$
\begin{equation}
   \sigma \rightarrow \sigma + v,
\label{shift}
\end{equation}
which can later be taken as the order parameter of the theory. After the spontaneous symmetry breaking, the Lagrangian for the LSMq is given by
\begin{align}
   \mathcal{L}&= \bar{\psi}(i\gamma^\mu \partial_\mu-M_q)\psi-g\bar{\psi}
   (\sigma+i\gamma_5\vec{\tau}\cdot \vec{\pi})\psi 
   - \frac{1}{2}M_\sigma^2\sigma^2\nonumber\\
   & - \frac{1}{2}M_\pi^2(\vec{\pi})^2
   -\frac{1}{2}(\partial_\mu\sigma)^2-
   \frac{1}{2}(\partial_\mu\vec{\pi})^2\nonumber\\
   &-\lambda v(\sigma^3+\sigma\vec{\pi}^2) -\frac{1}{4}\lambda (\sigma^4+2
   \sigma^2\vec{\pi}^2+\vec{\pi}^4)\nonumber\\
   & +\frac{a^2}{2}v^2-\frac{\lambda}{4}v^4. 
 \label{lagshift}
\end{align}
The shift in the $\sigma$-field produces that the quarks, the sigma boson and the three pions  acquire dynamical masses given by
\begin{align}
 M_q&=gv, \nonumber \\
 M_\sigma^2&=3\lambda v^2-a^2, \nonumber \\
 M_\pi^2&=\lambda v^2-a^2,
 \label{masses}
\end{align}
respectively. The tree-level potential is given by
\begin{equation}
    V^{\text{tree}}(v)=-\frac{a^2}{2}v^2+\frac{\lambda}{4}v^4,
    \label{treelevel}
\end{equation}
whose minimum is given by
\begin{equation}
    v_0=\sqrt{\frac{a^2}{\lambda}}.
\end{equation}
Since $v_0\neq 0$, we notice that the symmetry is spontaneously broken. 

The coupling constants $\lambda$ and $g$, as well as the mass parameter $a$ are to be determined from physical conditions valid at the phase transition. The minimum of the effective potential $v$ represents the order parameter that evolves when the system approaches chiral symmetry restoration at finite $T$ and/or $\mu_B$ and vanishes in the symmetric phase. In order to determine the transition conditions as function of temperature and quark chemical potential, we study the behavior of the effective potential.

The effective potential is computed beyond the mean field approximation. This means that we include, in addition to the tree-level contribution and the one-loop correction, both for bosons and fermions, also the ring diagrams contribution and the effective coupling constants. This correction accounts for the plasma screening effects~\cite{D&J}. Also, in order to consider a non-vanishing pion mass, we add to the Lagrangian an explicit symmetry breaking term and thus
\begin{eqnarray}
 \mathcal{L}\to\mathcal{L}'=\mathcal{L}+\frac{m_\pi^2}{2} v(\sigma+v),
\end{eqnarray}
and take the vacuum pion mass as $m_\pi\simeq 139$ MeV and the vacuum sigma mass $m_{\sigma}=500$ MeV. As a consequence of the explicit symmetry breaking, the effective potential at tree-level has a minimum at a value of $v$ given by
\begin{eqnarray}
    v_0=\sqrt{\frac{a^2+m_\pi^2}{\lambda}}.
\end{eqnarray}
The one-loop contribution contains vacuum as well as matter terms. The vacuum piece can potentially distort the tree-level potential, shifting the position of the minimum and distorting its curvature. Since the properties of vacuum should be independent of the perturbative order of its description, we introduce vacuum stability countertemrs $\delta a^2$ and $\delta\lambda$ so that the tree-level potential is now written as 
\begin{eqnarray}
    V^{\text{tree}}=-\frac{(a^2+m_\pi^2+\delta a^2)}{2}v^2+\frac{(\lambda+\delta \lambda)}{4}v^4.
    \label{newtree}
\end{eqnarray}
$\delta a^2$ and $\delta\lambda$ are computed from requiring that the minimum and the curvature at the minimum of the sum of the vacuum piece coming from the one-loop effective potential and the tree-level potential in Eq.~(\ref{newtree}) do not change with respect to the values they had without loop corrections. The explicit calculations of the counterterms is given in Appendix~A. These conditions yield
\begin{align}
    \delta a^2&=\frac{3}{16 \pi^2 \lambda } \Bigg [8 a^2 g^4-2 \gamma_E  a^2 \lambda^2+a^2 \lambda^2 \ln \left(\frac{4 \pi \mu_c^2}{m_\pi^2}\right)\nonumber \\
    &+a^2 \lambda^2 \ln \left(\frac{4 \pi \mu_c^2}{3 \left(a^2+m_\pi^2\right)-a^2}\right)+8 g^4 m_\pi^2-4 \lambda^2 m_\pi^2\Bigg]
    \nonumber\\
    \delta \lambda&=\frac{3}{16 \pi ^2}\Bigg[-8 g^4 \ln \left(\frac{4 \pi  \lambda \mu_c^2}{g^2 \left(a^2+m_\pi ^2\right)}\right)\nonumber \\
    &+3 \lambda ^2 \ln \left(\frac{4 \pi  \mu_c^2}{3 \left(a^2+m_\pi^2\right)-a^2}\right)+8 \gamma_E  g^4 \nonumber \\
    &-8 g^4-4 \gamma_E  \lambda ^2+4 \lambda ^2+\lambda ^2 \ln \left(\frac{4 \pi \mu_c^2}{m_\pi^2}\right)\Bigg].
    \label{deltalambda}
\end{align}

In order to find the ring diagrams correction to the effective potential, one needs to compute the meson self-energies, which serve as the temperature and density infrared regulators. Furthermore, an improved description of the properties of the phase diagram can be achieved when including temperature and density corrections to the couplings, which we carry up to one-loop order. 

The effective potential up to the ring diagrams contribution, the boson and fermion self-energies, and the effective couplings, can be analytically computed in the low and the high-temperature expansions. In the former, one considers that the largest energy scale is provided by $\mu_q$, such that $M/\mu_q$, $T/\mu_q\ll 1$, where $M$ is any of the particle masses. In the latter, it is only necessary to consider that $M/T\ll 1$, regardless of the relation between $T$ and $\mu_q$. The explicit expression for the effective potential in the low-$T$ expansion is given by
\begin{eqnarray}
V_{\text{LT}}^{\text{eff}}(v)&=&-\frac{a^2+m_{\pi}^{2}+\delta a^{2}}{2}v^2+\frac{\lambda+\delta\lambda}{4}v^4\nonumber \\
&-&3\Bigg\{\frac{ (M_\pi^2+\Pi_b^{\text{LT}})^2}{64\pi^2}\left[\ln\Big(\frac{4\pi \mu_c^2}{M_\pi^2+\Pi_b^{\text{LT}}}\Big)+\frac{5}{2}-\gamma_E\right]\nonumber \\
&+&T\bigg(\frac{T\sqrt{M_\pi^2+\Pi_b^{\text{LT}}}}{2\pi}\bigg)^{3/2}\text{Li}_{5/2}\Big(e^{-\sqrt{M_\pi^2+\Pi_b^{\text{LT}}}/T}\Big)\Bigg\} \nonumber \\
&-&\Bigg\{\frac{(M_\sigma^2+\Pi_b^{\text{LT}})^2}{64\pi^2}\left[\ln\left(\frac{4\pi \mu_c^2}{M_\sigma^2+\Pi_b^{\text{LT}}}\right)+\frac{5}{2}-\gamma_E\right]\nonumber \\
&+&T\bigg(\frac{T\sqrt{M_\sigma^2+\Pi_b^{\text{LT}}}}{2\pi}\bigg)^{3/2}\text{Li}_{5/2}\Big(e^{-\sqrt{M_\sigma^2+\Pi_b^{\text{LT}}}/T}\Big)\Bigg\} \nonumber \\
&+& N_c N_f\left\{ \frac{M_q^4}{16\pi^2}\left[\ln \left( \frac{4\pi \mu_c^2}{\left(\mu_q+\sqrt{\mu_q^2-M_q^2}\right)^2}\,\, \right)\right.\right. \nonumber \\
&-&\left.\gamma_E+\frac{5}{2}\right]-\frac{\mu_q\sqrt{\mu_q^2-M_q^2}}{24\pi^2}(2\mu_q^2-5M_q^2)\nonumber\\
&-&\frac{T^2\mu_q}{6}\sqrt{\mu_q^2-M_q^2}-\frac{7\pi^2T^4\mu_q}{360}\frac{(2\mu_q^2-3M_q^2)}{(\mu_q^2-M_q^2)^{3/2}}\nonumber \\
&+&\frac{31\pi^4 \mu_q M_q^4 T^6}{1008(\mu_q^2-M_q^2)^{7/2}}\Bigg\},
\label{VLT}
\end{eqnarray}
where for the matter contribution to the one-loop effective potential, we have included terms up to ${\mathcal{O}}(T)^6$. We use the expansion technique described in Ref.~\cite{chilenos} for the fermion case. Also, we have adopted the MS regularization scheme using $\mu_c e^{1/2}$ as the renormalization scale, with $\mu_c$ the critical quark chemical potential at $T=0$ in the QCD phase diagram. We have made tests scaling the renormalization scale $\mu_ce^{1/2}\to 3\mu_ce^{1/2}$ and found that the results vary only slightly. 

In the high-$T$ expansion, the effective potential is given by
\begin{eqnarray}
V_{\text{HT}}^{\text{eff}}(v)&=&-\frac{a^{2}+m_{\pi}^{2}+\delta a^{2}}{2}v^2+\frac{\lambda+\delta\lambda}{4}v^4\nonumber \\
&-&3\Bigg\{\frac{M_\pi^4}{64\pi^2}\bigg[\ln \Big(\frac{\mu_c^2}{4\pi T^2}\Big)+\frac{5}{2}-\gamma_E\bigg]+\frac{\pi^2 T^4}{90}\nonumber \\
&-&\frac{T^2 M_\pi^2}{24}+\frac{T(M_\pi^2+\Pi_b^{\text{HT}})^{3/2}}{12\pi}+\frac{\zeta(3)M_\pi^6}{96\pi^4 T^2}\Bigg \}\nonumber \\
&-&\Bigg\{ \frac{M_\sigma^4}{64\pi^2}\bigg[\ln \Big(\frac{\mu_c^2}{4\pi T^2}\Big)+\frac{5}{2}-\gamma_E\bigg]+\frac{\pi^2 T^4}{90}\nonumber \\
&-&\frac{T^2 M_\sigma^2}{24}+\frac{T(M_\sigma^2+\Pi_b^{\text{HT}})^{3/2}}{12\pi}+\frac{\zeta(3)M_\sigma^6}{96\pi^4 T^2} \Bigg\}\nonumber \\
&+&\frac{N_c N_f}{16\pi^2}\Bigg\{ M_q^4 \bigg[ \ln \Big(\frac{\mu_c^2}{4\pi T^2} \Big)+\frac{5}{2}-\gamma_E\nonumber \\
&-&\psi^{(0)}\Big(\frac{1}{2}+\frac{\mathrm{i}\mu_q}{2\pi T}\Big)-\psi^{(0)}\Big(\frac{1}{2}-\frac{\mathrm{i}\mu_q}{2\pi T}\Big) \bigg]\nonumber \\
&-&8M_q^2T^2\big[\text{Li}_2(-e^{\mu_q/T})+\text{Li}_2(-e^{-\mu_q/T})\big]\nonumber \\
&+&32T^4\big[\text{Li}_4(-e^{\mu_q/T})+\text{Li}_4(-e^{-\mu_q/T})\big]\nonumber \\
&+&\frac{M_q^6}{6T^2}\left[\psi^{(2)}\Big(\frac{3}{2}+\frac{\mathrm{i}\mu_q}{2\pi T}\Big)\right.\nonumber \\
&+&\left.\psi^{(2)}\Big(\frac{3}{2}-\frac{\mathrm{i}\mu_q}{2\pi T}\Big)\right]\Bigg\},
\label{VHT}
\end{eqnarray}
where for the matter contribution we have included terms up to ${\mathcal{O}}(M)^6$, using the expansion technique described in Ref.~\cite{D&J}.

The expression for the boson self-energies, $\Pi_b$, in the low- and high-temperature approximations are given by
\begin{equation}
    \Pi_b^{\text{LT}}=N_c N_f g^2\bigg(\frac{\mu_q^2}{2\pi^2}+\frac{T^2}{6}\bigg)
    \label{PiLT}
\end{equation}
and
\begin{eqnarray}
    \Pi_b^{\text{HT}}&=&\frac{\lambda T^2}{2}-N_c N_f \frac{g^2 T^2}{\pi^2}\nonumber\\
    &\times&\left[\text{Li}_2(-e^{\mu_q/T})+\text{Li}_2(-e^{-\mu_q/T})\right],
    \label{PiHT}
\end{eqnarray}
respectively. 

The temperature and density corrections to the couplings, accounting for the modification of the intensity of the interaction around the phase transition region, are explicitly computed in Appendices B and C,  and given by  
\begin{align}
    \lambda^{\text{eff}}_{HT}=\lambda& \Bigg[1-\frac{24 \lambda}{4}\frac{14}{4\pi^{2}}\bigg[\frac{\pi T}{2(\Pi_b^{HT})^{1/2}}\nonumber \\
    &+\frac{1}{2}\ln\bigg(\frac{(\Pi_b^{HT})^{1/2}}{4\pi T}\bigg)+\frac{1}{2}\gamma_{E}\bigg] \Bigg]
    \label{lambdaeffLT}
\end{align}

\begin{align}
    \lambda^{\text{eff}}_{LT}=\lambda& \Bigg[1+\frac{24 \lambda}{4}14\bigg(\frac{T^{3}}{8\pi^3}\bigg)^{1/2}\bigg[\frac{\text{Li}_{3/2}(e^{-\frac{(\Pi_b^{LT})^{1/2}}{T}})}{4(\Pi_b^{LT})^{3/4}}\nonumber \\
    &-\frac{\text{Li}_{1/2}(e^{-\frac{(\Pi_b^{LT})^{1/2}}{T}})}{2T (\Pi_b^{LT})^{1/4}}\bigg]\Bigg]
    \label{lambdaeffHT}
\end{align}

\begin{align}
    &g^{\text{eff}}_{LT}=g\Bigg[1+4g^2\bigg[\frac{\sqrt{\mu_q^{2}-m_f^2}}{8\pi^{2}\mu_q}\nonumber \\
    &-\frac{1}{8\pi^{2}}\ln\bigg(\frac{\mu_q+\sqrt{\mu_q^{2}-m_f^2}}{m_f}\bigg)\nonumber \\
    &+
    \frac{2(\Pi_b^{LT})^2}{16\pi^{2}(m_{f}^{2}-\Pi_b^{LT})^{2}}\bigg[ \ln\bigg(\frac{\mu_q+\sqrt{\mu_q^{2}-m_{f}^{2}}}{m_{f}}\bigg)\nonumber \\
    &-\bigg(1-\frac{m_f^2}{\mu_q^{2}}\bigg)^{1/2}(1+\frac{\mu_q^{2}-m_{f}^{2}}{\Pi_b^{LT}})\bigg]+\frac{T^{2}}{48}\nonumber \\
    &\times
    \frac{(-2m_{f}^{6}+m_{f}^{4}(\mu_q^2+2\Pi_b^{LT})-\mu_q^2m_f^{2}\Pi_b^{LT}-2\mu_q^4\Pi_b^{LT})}{\mu_q^3\sqrt{\mu_q^2-m_{f}^{2}}(m_{f}^{2}-\Pi_b^{LT})^{2}}\bigg] \Bigg].
    \label{geffLTtt}
\end{align}
and
\begin{align}
    g^{\text{eff}}_{HT}=g&\Bigg[1-4g^2\frac{T^2}{4\pi^{2} \Pi_b^{HT}}\Big[\frac{\pi(\Pi_b^{HT})^{1/2} }{T}\nonumber \\ &+Li_{2}(-e^{\frac{-\mu_{q}}{T}})+Li_{2}(-e^{\frac{\mu_{q}}{T}})\Big]\Bigg],
    \label{geffHT}
\end{align}
where for the fermion mass $m_{f}$ in the low temperature approximation of $g^{eff}_{LT}$ we have used the fermion self-energy given by
\begin{align}
    \Pi_{LT}^{f}=g^{2}\bigg(\frac{\mu^{2}}{2\pi^{2}}+\frac{T^{2}}{2}\bigg).
    \label{ec:piFermionLT}
\end{align}
These effective couplings enter the effective potential through the boson and fermion masses which are now written as
\begin{align}
    M_q&=g^{eff}v, \nonumber \\
 M_\sigma^2&=3\lambda^{eff} v^2-a^2, \nonumber \\
 M_\pi^2&=\lambda^{eff} v^2-a^2,
 \label{newmasses}
\end{align}
as well as through the counterterms $\delta a^2$ and $\delta\lambda$.

Before proceeding to explore the properties of the phase transition including the superstatistics effects, we first
need to determine the free parameters of the LSMq which are appropriate to the conditions of finite temperature and density around the transition lines. 

\section{Determination of the free parameters}\label{sec4}

\begin{figure}[t]
\begin{center}
\includegraphics[scale=0.58]{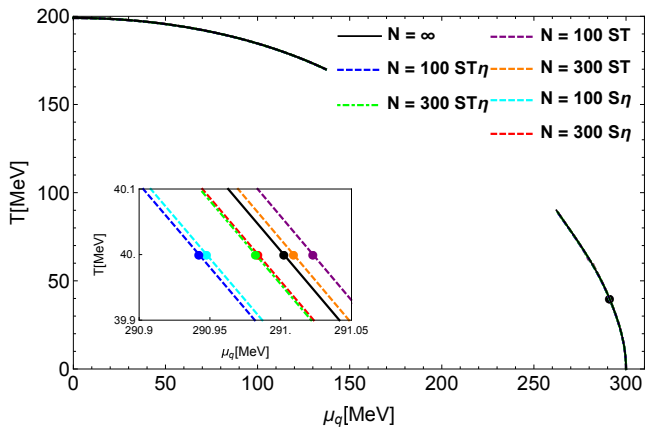}
\end{center}
\caption{QCD phase diagram obtained from the low- and high-temperature approximations with $\mu_c=300$ MeV and $\lambda=1.5$, and the corresponding $g=1.77$ and $a=102$ MeV. The CEP is located within the full circles on each curve obtained using the low temperature expansion.}
\label{fig1}
\end{figure}

The effective potential contains three free parameters that need to be fixed, namely the couplings $\lambda$ and $g$, and the mass parameter $a$. These parameters are to be determined using physical input valid at the phase transition. For this purpose, we enforce that at $T=0$ and for a critical value of the quark chemical potential $\mu_c=\mu_c^B/3$~\cite{models}, with $\mu_c^B$ the critical baryon chemical potential, the effective potential describes a first order phase transition. In analogy with the Hagedorn's limiting temperature concept~\cite{Hagedorn} extended to finite $\mu_B$, we take $\mu_B\simeq m_B$, where $m_B\simeq$ 1 GeV is the typical value of the baryon mass. Recall that for a first order phase transition, the effective potential develops two degenerate minima. The value of $v\equiv v^*\neq 0$ at the minimum becomes a new quantity that needs to be also determined. However, as we proceed to show, the conditions to describe a first order phase transition provide only three equations to determine the four unknowns. Our strategy to find the solutions consists of finding the model parameters when varying one of them and choose $\lambda$ as the parameter to vary.
First, in order to determine $a$, we use the relation between the $\sigma$ and $\pi$ dynamical masses at $T=0$ and $\mu_q=\mu_c$. This involves  Eqs.~(\ref{masses}), including the matter corrections coming from the self-energies, Eq.~(\ref{PiLT}). Thus, we have
\begin{equation}
    a=\Bigg(\frac{M_\sigma^2-3M_\pi^2-2\Pi_b^{\text{LT}}\Big|_{T=0,\mu_q=\mu_c}}{2}\Bigg)^{1/2}.
    \label{fixingA}
\end{equation}
The remaining two equations for the two unknowns, $v^*$ and $g$, are given by
\begin{align}
    \frac{\partial V_{\text{LT}}^{\text{eff}}(T=0,\mu_q=\mu_c)}{\partial v}\Bigg |_{v=v^*}&=0, \nonumber \\
    V_{\text{LT}}^{\text{eff}}(T=0,\mu_q=\mu_c)\Bigg |_{v=0}&=V_{\text{LT}}^{\text{eff}}(T=0,\mu_q=\mu_c)\Bigg |_{v=v^*}.
    \label{seteqpara}
\end{align}
Equations~(\ref{seteqpara}) describe the conditions for the effective potential to show degenerate minima at $T=0$ for $\mu_q=\mu_c$, one of them at $v=0$ and the second one at $v=v^*$. The appropriate expression to be used for the self-energy corresponds to the case of the low temperature regime.

Although the above described procedure fixes the model parameters for conditions valid at the putative first order phase transition, we also require consistency of the description with the crossover transition at $\mu_B=0$ and finite $T$. From LQCD~\cite{latticeTc} calculations, it is well-known that this transition happens at $T^c_0\simeq 155$ MeV for 2+1 light flavors and at $T^c_0\simeq 170$ MeV for 2 light flavors. In order to get solutions of Eqs.~(\ref{fixingA}) and~(\ref{seteqpara}) satisfying the conditions at both ends of the transition curve with $T_c(\mu=0)<200$ MeV we find that $\lambda$ is required to lie between the very restricted range $1.4<\lambda <1.5$. The lowest limit for these $ \lambda$ values corresponds to a choice for $\mu_c=300$ MeV, whereas the upper limit corresponds to $\mu_c=290$ MeV. 

\section{QCD phase diagram}\label{sec5}

\begin{figure}[t]
\begin{center}
\includegraphics[scale=0.58]{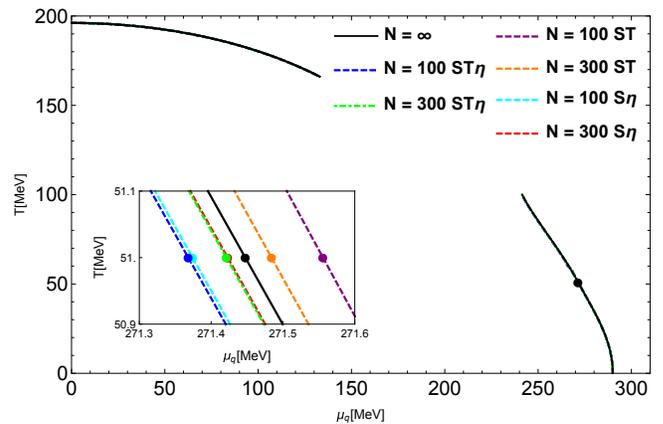}
\end{center}
\caption{QCD phase diagram obtained from the low- and high-temperature approximations with $\mu_c=290$ MeV and $\lambda=1.6$, and the corresponding $g=1.85$ and $a=93.4$ MeV. The CEP is located within the full circles on each curve obtained using the low temperature expansion.}
\label{fig2}
\end{figure}

We now, proceed to use the effective potential we have determined, into the supersatistics formalism. Following the procedure developed in Sec.~\ref{sec2}, we substitute either Eq.~(\ref{VLT}) or Eq.~(\ref{VHT}) into Eq.~(\ref{ec:partitionfunction0}) for the low and high-temperature descriptions, respectively. In this manner, we can obtain the partition function using Eq.~(\ref{ec:boxedz0}) and from its logarithm, we write the effective potential, including the superstatistics corrections, as
\begin{equation}
    V^{\text{eff}}_{\text{sup}}=-\frac{1}{\Omega\beta}\ln[Z].
    \label{effpotsup}
\end{equation}
To explore the QCD phase diagram, we identify the transition lines where chiral symmetry is restored. The procedure consists of finding the transition temperatures and chemical potentials using first the high temperature approximation for the effective potential. We  start from $\mu_q=0$ and stop when the ratio of the critical chemical potential and temperature is $\approx 0.8$ from where we start using the low-temperature approximation for the effective potential to continue finding the critical curve.

From this procedure, we find either second or first order transitions. For the former, the vacuum expectation value continuously moves from a finite value to zero and the phase transition occurs when this vanishes. For the latter, the phase transition is identified when the two developed minima become degenerate. 

Figure~\ref{fig1} shows the effective QCD phase diagram using $\lambda=1.5$ and $\mu_c=300$ MeV with $a=102$ MeV  and $g=1.77$. In the region where the HT approximation is valid, we find only second order phase transitions (our proxy for the crossover phase transition) and these are shown as the curve with $0\leq \mu_q < 150$ MeV. In contrast, both second and first order phase transitions are found deep in the phase diagram where the LT approximation is valid. This happens for $260 \ \text{MeV}< \mu \leq \mu_c$. The CEP location, where the phase transition changes order, is indicated with a full circle at $\mu_q \approx 291$ MeV and $T \approx 40$ MeV. Figure~\ref{fig1}, also shows the variation on the transition line and the CEP due to the superstatistics effects in temperature (S$_T$), in $\mu_q/T$ (S$_\eta$) and in both parameters at the same time (S$_{T\eta}$), with $N=100$ and $N=300$ for all combinations. We notice that as $N$ increases, the corresponding CEP moves towards the corresponding CEP in the thermodynamic limit, labeled as $N=\infty$. The changes due to any of the superstatistics effects are quite small and for all practical purposes they are neglegible. Figure~\ref{fig2}, shows the effective QCD phase diagram using other set of allowed parameters, $\lambda=1.6$ and $\mu_c=290$ MeV with, $a=93.4$ MeV and $g=1.85$, the general features of the phase diagram in this case do not change with respect to the previous case. However, it is relevant to mention that the CEP is now located at $\mu_q \approx 271$ MeV and $T \approx 51$ MeV. Nevertheless, the superstatistics effects are also negligible. Finally, we notice that the systematics of the CEP displacement for the two explored sets of parameters is the same.



\section{Summary and Conclusions}\label{sec6}
In this work we have used the LSMq to locate the CEP in the effective QCD phase diagram taking into account fluctuations in the temperature and the quark chemical potential. For this purpose, we have implemented the superstatistics scenario assuming that both fluctuations in the temperature and quark chemical potential are described according to a $\chi^{2}$ distribution. We computed the superstatistics effective potential up to order $1/N$. To numerically find the effects, $N$ has been taken as about half the number of nucleons in the collision of heavy ions.

In order to locate the CEP we have used the LSMq, computing the effective potential in the high- and low- temperature approximations up to sixth order, including the ring diagram contribution which accounts for the plasma screening effects. We fix the model parameters, namely the coupling constants and the mass parameter, imposing a first order phase transition at zero temperature and a finite critical quark chemical potential, using the Hagedorn limit temperature concept applied to finite baryon density. We fix this critical baryonic chemical potential $\mu_{B}^{c}\sim 1$ GeV.

We find that the CEP displacement due to fluctuations in temperature and chemical potential are of order $\sim 0.1$ MeV and are much smaller compared with the case when only fluctuations in temperature are considered, and the model parameters are fixed in the high-temperature effective potential~\cite{AHHLZ}. As noted also in Ref.~\cite{AHHLZ}, $N$ can be associated with the number of participants in the heavy-ion collision, the specific heat and the smallest of the mass numbers of the colliding nuclei. Thus, in order to give a more accurate estimation of how much the CEP is displaced under appropriate experimental conditions, these need to be included for the estimation of the parameter $N$. 

\section*{ACKNOWLEDGEMENTS}

Support for this work has been received in part by UNAM-DGAPA-PAPIIT grant number AG100219. L. A. H. and V. K.-P. acknowledge support from PAPIIT-DGAPA-UNAM fellowships. R. Zamora acknowledges support from FONDECYT (Chile) under grant No. 1200483. S.H.O. acknowledges support from the U.S. DOE under Grant No. DE-FG02-00ER41132 and the Simons Foundation under the Multifarious Minds Program Grant No. 557037.

\section*{Appendices}

\subsection{Vacuum stability conditions}\label{App1}
\renewcommand{\theequation}{A\arabic{equation}}
\setcounter{equation}{0}
The vacuum stability conditions are introduced to ensure that $v_0$ and the sigma-mass maintain their tree level values, even after including the vacuum pieces stemming from the one-loop corrections. These conditions are
\begin{align}\label{v02}
\frac{1}{2v}\frac{dV^{\text{vac}}}{dv} \Big|_{v=v_0}&=0,\nonumber \\
\frac{d^2V^{\text{vac}}}{dv^2} \Big|_{v=v_0}&=2a^2+2m_\pi^2,
\end{align}
where $V^{\text{vac}}$ is the one-loop vacuum piece of the effective potential. The solution for the counterterms $\delta a^2$ and $\delta \lambda$ is given by 
\begin{align}
    \delta a^2&=\frac{3}{16 \pi^2 \lambda } \Bigg [8 a^2 g^4-2 \gamma_E  a^2 \lambda^2\nonumber \\
    &+a^2 \lambda^2 \ln \left(\frac{4 \pi \mu_c^2}{m_\pi^2}\right)+a^2 \lambda^2 \ln \left(\frac{4 \pi \mu_c^2}{3 \left(a^2+m_\pi^2\right)-a^2}\right)\nonumber \\
    &+8 g^4 m_\pi^2-4 \lambda^2 m_\pi^2\Bigg]
    \label{deltaa2}
\end{align}
and
\begin{align}
    \delta \lambda&=\frac{3}{16 \pi ^2}\Bigg[-8 g^4 \ln \left(\frac{4 \pi  \lambda \mu_c^2}{g^2 \left(a^2+m_\pi ^2\right)}\right)\nonumber \\
    &+3 \lambda ^2 \ln \left(\frac{4 \pi  \mu_c^2}{3 \left(a^2+m_\pi^2\right)-a^2}\right)+8 \gamma_E  g^4 \nonumber \\
    &-8 g^4-4 \gamma_E  \lambda ^2+4 \lambda ^2+\lambda ^2 \ln \left(\frac{4 \pi \mu_c^2}{m_\pi^2}\right)\Bigg].
    \label{deltalambda1}
\end{align}

\begin{widetext}
\begin{figure*}[t]
\begin{center}
\includegraphics{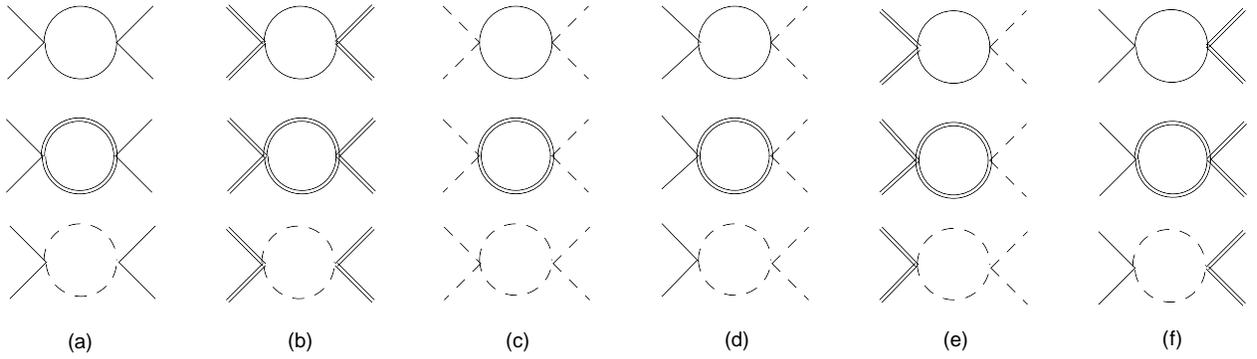}
\end{center}
\caption{One-loop Feynman diagrams that contribute to the thermal correction of the coupling $\lambda$. The dashed line denotes the charged pion, the continuous line is the sigma and the double line represents
the neutral pion.}
\label{figlambdacorrection}
\end{figure*}
\end{widetext}

\subsection{Effective coupling constants ($\lambda$)}\label{App2}
\renewcommand{\theequation}{B\arabic{equation}}
\setcounter{equation}{0}

We now compute the one-loop correction to the coupling $\lambda$, including thermal effects in the high temperature and low temperature regimes. Figure~\ref{figlambdacorrection} shows the Feynman diagrams that contribute to this correction. Columns (a), (b), (c), (d), (e) and (f) contribute to the correction to the $\sigma^4$, $(\pi^0)^4$, $(\pi^+)^2(\pi^-)^2$, $\sigma^2\pi^+\pi^-$, $(\pi^0)^2\pi^+\pi^-$ and  $\sigma^2(\pi^0)^2$ terms of the interaction Lagrangian in Eq.~(\ref{lagrangian}), respectively. Since each of these corrections lead to the same result, we concentrate on the diagrams in column (a). Each of the three diagrams involves two propagators of the same boson. For the first two diagrams the intermediate bosons are neutral and for the third one the intermediate bosons are charged. Therefore the expression for the diagrams can be obtained from 
\begin{equation}
 I(P_i;m_i^2)=T\sum_n\int\frac{d^3k}{(2\pi)^3}D(P_i-K)D(K),
\label{general}
\end{equation}
where $P_i\equiv(\omega,{\mbox{\bf{p}}})$ is the total incoming four-momentum, $K\equiv (\omega_n,{\mbox{\bf{k}}})$, $\omega_n = 2n\pi T$ is the bosonic Matsubara frequency and $D$ is the boson propagators defined as 
\begin{equation}
    D(K)=\frac{i}{K^2-m^2}.
    \label{bosonicop}
\end{equation}
\begin{figure}[H]
\begin{center}
\includegraphics[scale=0.58]{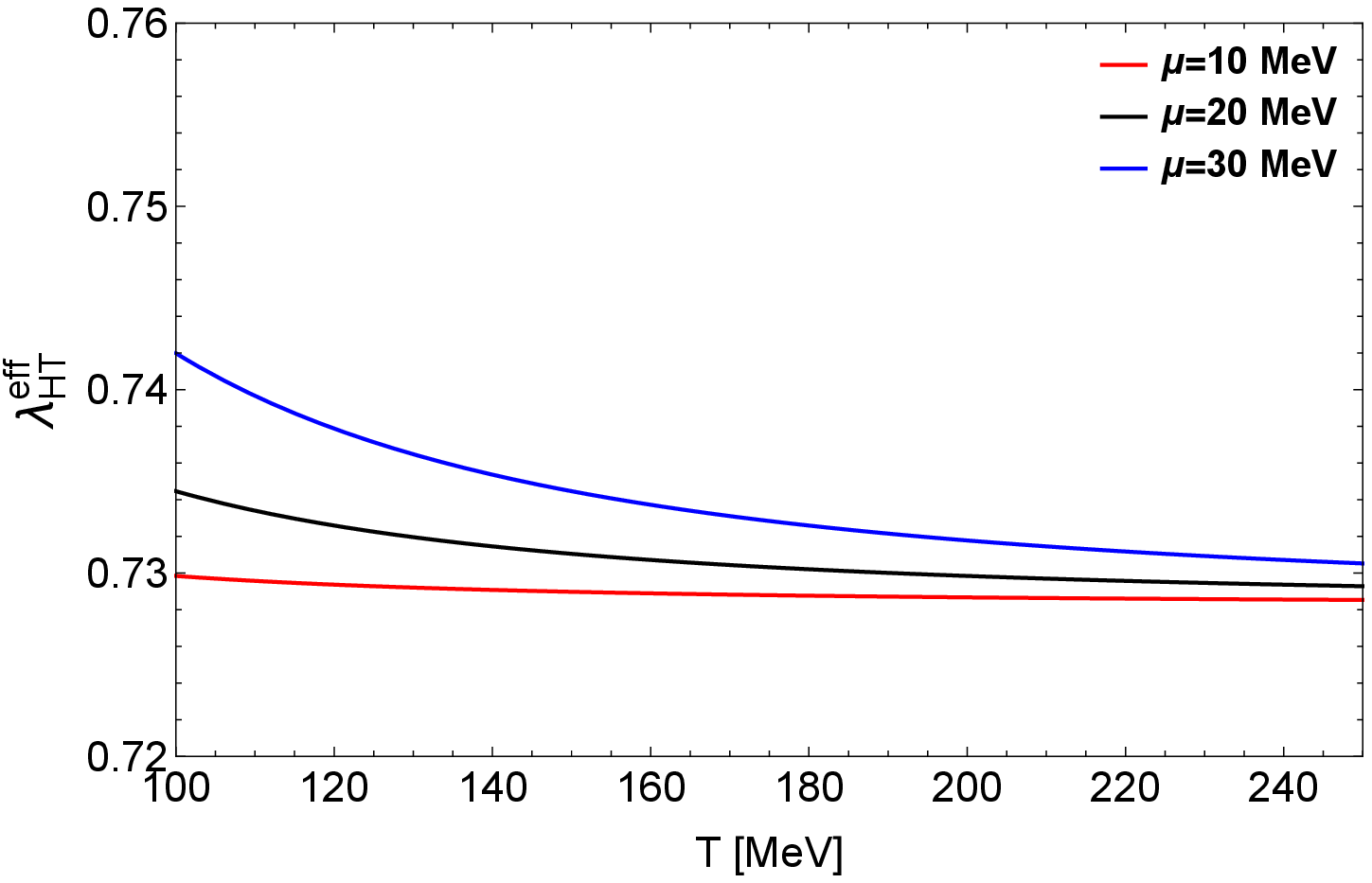}
\end{center}
\caption{Effective boson self-coupling $\lambda_{\text{eff}}$, in the high-temperature approximation, as a function of $T$.}
\label{fig6}
\end{figure}

Considering the permutation factors and the contribution from the $s$, $t$ and $u$-channels, the correction to the self-coupling $\lambda$ to one-loop order is given by
\begin{equation}
   \lambda_{\mbox{\small{eff}}}=\lambda
   \left[1 + \frac{24\lambda}{4} \times 14I(P_i;m^2)
   \right],
\label{lambdaeff}
\end{equation}
we have to compute Eq.~(\ref{general}), explicitly
\begin{eqnarray}
    I(P_i;m^2)&=&-T \int \frac{d^3k}{(2\pi)^3} \sum_n \frac{1}{[\omega_n^2+\textbf{k}^2+m^2]} \nonumber \\
    &\times&\frac{1}{[(\omega_n-\omega)^2+(\textbf{k}-\textbf{p})^2+m^2]}. \label{sum1}
\end{eqnarray}
The sum in Eq.~(\ref{sum1}) is calculated using Ref.~\cite{LeBellac}. We obtain 
\begin{eqnarray}
   I(P_i;m^2)&=&  - \int \frac{d^3k}{(2\pi)^3} \nonumber \\
   &\times&\sum_{s1,s2}\frac{-s_1 s_2}{4E_1 E_2} \frac{(1+f(s_1 E_1)+f(s_2 E_2))}{ i \omega - s_1 E_1 - s_2 E_2}, \nonumber \\
\end{eqnarray}
where 
\begin{eqnarray}
    E_1&=&\sqrt{\textbf{k}^2-\textbf{p}^2+m^2} \nonumber \\
    E_2&=&\sqrt{\textbf{k}^2+m^2} \nonumber \\
    f(x)&=& \frac{1}{e^{x/T}-1}.
\end{eqnarray}
Calculating in the infrared limit ($E_1=E_2\equiv E$), we get
\begin{eqnarray}
  I(0;m^2)&=&-\int \frac{d^3k}{(2\pi)^3} \frac{1}{8 E^3} (1+4 f(E)) \nonumber \\
 &=&-\frac{1}{16 \pi^2} \int_{0}^{\infty} \frac{dk k^2}{(k^2+m^2)^{3/2}} \nonumber \\
 &\times& \left(1 + \frac{4}{e^{\sqrt{k^2+m^2}/T}-1} \right). \nonumber \\ \label{lambda1}
\end{eqnarray}


First, we study the high temperature (HT) case. We focus on the  matter term of Eq.~(\ref{lambda1}),
\begin{eqnarray}
   I(0;m^2)&=& -\frac{1}{4 \pi^2} \int_{0}^{\infty} \frac{dk k^2}{(k^2+m^2)^{3/2}} \frac{1}{e^{\sqrt{k^2+m^2}/T}-1} \nonumber \\
I(0;m^2)_{\text{HT}}&=& \frac{1}{2\pi^2} \nonumber \\
&\times&\frac{\partial}{\partial m^2}  \int_{0}^{\infty} \frac{dkk^2}{(k^2+m^2)^{1/2}} \frac{1}{e^{\sqrt{k^2+m^2}/T}-1},\nonumber \\ \label{ltt1}
\end{eqnarray}
notice that in the second of Eqs.~(\ref{ltt1}) we retain only the dominant term. We define $y \equiv m/T$ and $x\equiv k/T$ and get
\begin{equation}
  I(0;m^2)_{\text{HT}}= \frac{1}{2\pi^2} \frac{\partial}{\partial y^2}\int_0^\infty \frac{dx x^2}{(x^2+y^2)^{1/2}} \frac{1}{e^{\sqrt{x^2+y^2}}-1},
\end{equation}
The integrals over $x$
can be expressed in terms of the well known functions \cite{Kapusta}
\begin{eqnarray}
  h_n(y)&=&\frac{1}{\Gamma (n)}\int_0^\infty 
   \frac{dx\ x^{n-1}}{\sqrt{x^2+y^2}}\frac{1}{e^{\sqrt{x^2+y^2}}-1}\nonumber \\
   f_n(y)&=&\frac{1}{\Gamma (n)}\int_0^\infty 
   \frac{dx\ x^{n-1}}{\sqrt{x^2+y^2}}\frac{1}{e^{\sqrt{x^2+y^2}}+1},
\label{handf}
\end{eqnarray}
which satisfy the differential equations
\begin{eqnarray}
   \frac{\partial h_{n+1}}{\partial y^2}&=&-\frac{h_{n-1}}{2n}\nonumber \\
   \frac{\partial f_{n+1}}{\partial y^2}&=&-\frac{f_{n-1}}{2n}.
\label{diffeq}
\end{eqnarray}
Therefore,
\begin{equation}
I(0;m^2)_{\text{HT}}= -\frac{1}{4 \pi^2} h_1(y).
\end{equation}
Using the high temperature expansions for $h_1(y)$ and $f_1(y)$~\cite{Kapusta}
\begin{eqnarray}
   h_1(y)&=&\frac{\pi}{2y} + \frac{1}{2}\ln\left(\frac{y}{4\pi}\right) + \frac{1}{2}\gamma_E + \ldots\nonumber \\
   f_1(y)&=&-\frac{1}{2}\ln\left(\frac{y}{\pi}\right) - \frac{1}{2}\gamma_E + \ldots,
\label{expansions}
\end{eqnarray}
and keeping the leading terms, we get
\begin{equation}
  I(0;m^2)_{\text{HT}}  =-\frac{1}{4 \pi^2} \biggl[ \frac{\pi T}{2 m}+ \frac{1}{2} \ln\left(\frac{m}{4\pi T}\right)+\frac{1}{2} \gamma_E \biggr]. \label{lambdahtt}
\end{equation}
Using Eq.~(\ref{lambdahtt}) into Eq.~(\ref{lambdaeff}), we finally obtain
\begin{align}
    \lambda^{\text{eff}}_{HT}=\lambda& \Bigg[1-\frac{24 \lambda}{4}\frac{14}{4\pi^{2}}\bigg[\frac{\pi T}{2(\Pi_b^{HT})^{1/2}}\nonumber \\
    &+\frac{1}{2}\ln\bigg(\frac{(\Pi_b^{HT})^{1/2}}{4\pi T}\bigg)+\frac{1}{2}\gamma_{E}\bigg] \Bigg].
    \label{lambdaeffLT1}
\end{align}
We now calculate the low temperature (LT) case. We focus on the  matter term of Eq.~(\ref{lambda1}) 
\begin{eqnarray}
   I(0;m^2)&=& -\frac{1}{4 \pi^2} \int_{0}^{\infty} \frac{dk k^2}{(k^2+m^2)^{3/2}} \frac{1}{e^{\sqrt{k^2+m^2}/T}-1} \nonumber \\
  I(0;m^2)_{\text{LT}} &=& \frac{1}{2 \pi^2}\nonumber \\
  &\times&\frac{\partial}{\partial m^2}  \int_{0}^{\infty} \frac{dkk^2}{(k^2+m^2)^{1/2}} \frac{1}{e^{\sqrt{k^2+m^2}/T}-1}.\nonumber \\\label{lttt2}
\end{eqnarray}
Note that in the second equality of Eq.~(\ref{lttt2}), there is a term that is zero when $T \rightarrow 0$, if we make the change of variable $w=\sqrt{k^{2}+m^{2}}$, then $dk=\frac{wdw}{\sqrt{w^{2}-m^{2}}}$. Therefore, we get 
\begin{equation}
     I(0;m^2)_{\text{LT}}=\frac{1}{2\pi^{2}}\frac{\partial}{\partial m^{2}}\int_{m}^{\infty}dw\sqrt{w^{2}-m^{2}}\frac{e^{-w/T}}{1-e^{-w/T}}.
    \label{ec:lambdaLTL1}
\end{equation}

\begin{figure}[t]
\begin{center}
\includegraphics[scale=0.58]{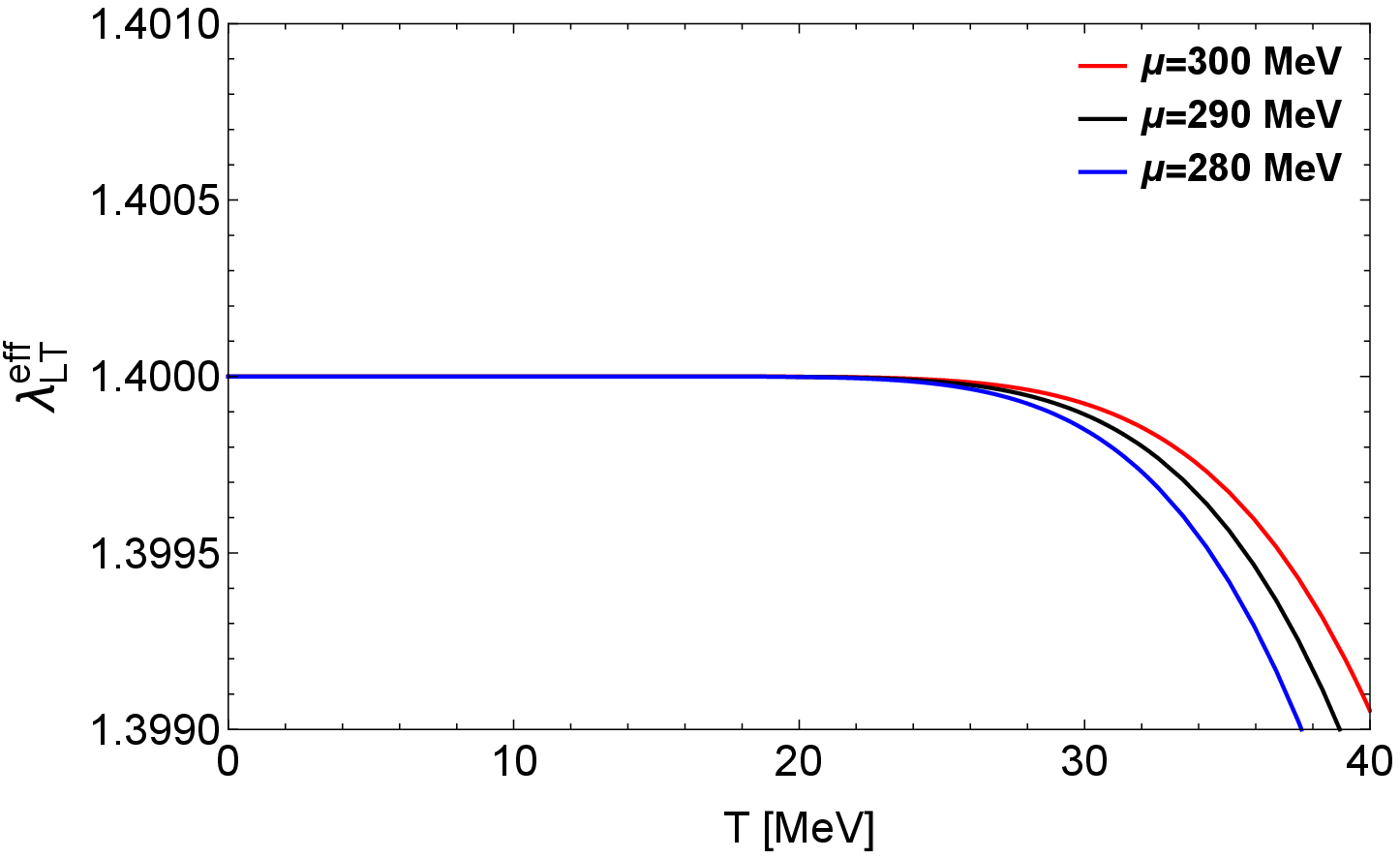}
\end{center}
\caption{Effective boson self-coupling $\lambda_{\text{eff}}$, in the low-temperature approximation, as a function of $T$.}
\label{fig5}
\end{figure}

as $T\rightarrow 0$, then $e^{-w/T}<<1$ and we can use the geometric series in the form 
\begin{equation}
    \frac{e^{-w/T}}{1-e^{-w/T}}= \sum_{n=1}^{\infty}e^{-nw/T},
    \label{ec:serie geometrica}
\end{equation}
thus, using Eq.~(\ref{ec:serie geometrica}) into Eq.~(\ref{ec:lambdaLTL1}), we obtain
\begin{equation}
   I(0;m^2)_{\text{LT}}=\frac{1}{2\pi^{2}}\frac{\partial}{\partial m^{2}}\sum_{n=1}^{\infty}\int_{0}^{\infty}dw\sqrt{w^{2}-m^{2}}e^{-\frac{nw}{T}}.
    \label{ec:lambdaLTL2}
\end{equation}
Notice that the integral of (\ref{ec:lambdaLTL2}) gives
\begin{eqnarray}
    \int_{0}^{\infty}dw\sqrt{w^{2}-m^{2}}e^{-\frac{nw}{T}}&=& \frac{mT}{n}K_{1}(\frac{mn}{T}) \nonumber \\
    &\approx&\bigg(\frac{T^{3}m\pi}{2n^{3}}\bigg)^{1/2}e^{-\frac{mn}{T}}, \nonumber \\
    \label{ec:bessel}
\end{eqnarray}
where in the last equality we have expanded in a Taylor series for $T<<1$. Therefore we get
\begin{eqnarray}
     I(0;m^2)_{\text{LT}}&=&\frac{1}{2\pi^{2}}\frac{\partial}{\partial m^{2}}\sum_{n=1}^{\infty}\bigg(\frac{T^{3}m\pi}{2 n^{3}}\bigg)^{1/2}e^{-\frac{m\pi}{T}} \nonumber \\
     &=&\frac{1}{2\pi^{2}}\frac{\partial}{\partial m^{2}}\bigg(\frac{T^{3}m\pi}{2}\bigg)^{1/2}\sum_{n=1}^{\infty}\frac{e^{-\frac{mn}{T}}}{n^{3/2}}\nonumber\\
    &=&\frac{1}{2\pi^{2}}\frac{\partial}{\partial m^{2}}\bigg(\frac{T^{3}m\pi}{2}\bigg)^{1/2}\text{Li}_{3/2}(e^{-m/T}),
    \label{ec:lambdaLTLE} \nonumber \\
\end{eqnarray}
and thus,
\begin{eqnarray}
    I(0;m^2)_{\text{LT}}&=&\frac{1}{2\pi^{2}}\bigg(\frac{T^{3}\pi}{2}\bigg)^{1/2}\bigg[\frac{1}{4m^{3/2}}Li_{3/2}(e^{-\frac{m}{T}})\nonumber \\
    &-&\frac{1}{2T m^{1/2}}Li_{1/2}(e^{-\frac{m}{T}})\bigg].
    \label{ec:lambdaLTL3}
\end{eqnarray}
Finally, using Eq.~(\ref{ec:lambdaLTL3}) into Eq.~(\ref{lambdaeff}), we get
\begin{align}
    \lambda^{\text{eff}}_{LT}=\lambda& \Bigg[1+\frac{24 \lambda}{4}14\bigg(\frac{T^{3}}{8\pi^3}\bigg)^{1/2}\bigg[\frac{\text{Li}_{3/2}(e^{-\frac{(\Pi_b^{LT})^{1/2}}{T}})}{4(\Pi_b^{LT})^{3/4}}\nonumber \\
    &-\frac{\text{Li}_{1/2}(e^{-\frac{(\Pi_b^{LT})^{1/2}}{T}})}{2T (\Pi_b^{LT})^{1/4}}\bigg]\Bigg].
    \label{lambdaeffHT1}
\end{align}

\subsection{Effective coupling constants (g)}\label{App3}
\renewcommand{\theequation}{C\arabic{equation}}
\setcounter{equation}{0}
\begin{figure}[t]
\begin{center}
\includegraphics[scale=0.5]{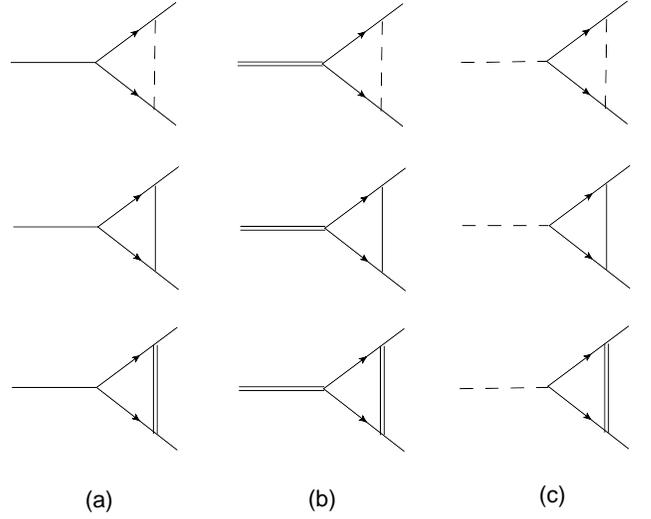}
\end{center}
\caption{One-loop Feynman diagrams that contribute to the thermal  correction of the coupling $g$. The dashed line denotes the charged pion, the continuous line is the sigma, the double line represents the neutral pion and the continuous line with arrows represents the quarks.}
\label{gcorrection}
\end{figure}

We now turn to the calculation for the one-loop correction of the coupling $g$. Figure~\ref{gcorrection} shows the Feynman diagrams that contribute to this correction. Columns (a), (b) and (c) contribute to the correction to the quark-$\sigma$, quark-$\pi^0$ and quark-$\pi^\pm$ terms of the interaction Lagrangian of Eq.~(\ref{lagrangian}), respectively. Since each of these corrections lead to the same result, we concentrate on the first diagrams in column (a).  The expression for this diagram is written as
\begin{eqnarray}
    \label{ec:feynman1}
    &&G(P_i,m^2)=\int \frac{d^{4}k}{(2\pi)^4}\bar{u}(P_2)(-ig)(i\tau_{3}\gamma_{5})S(P_1-K)\nonumber \\
    &\times&(-ig)S(P_2-K)(-ig)(i\tau_{3}\gamma_{5})D(K)u(P_1),
\end{eqnarray}

\begin{figure}[t]
\begin{center}
\includegraphics[scale=0.5]{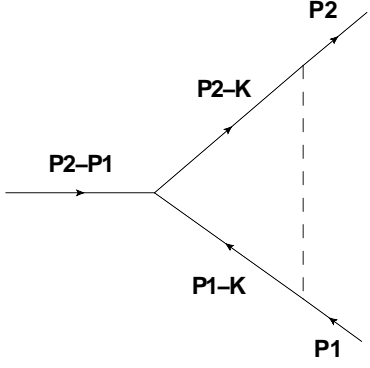}
\end{center}
\caption{Feynman diagram defining the momentum corresponding to each line for a generic diagram in Fig.~6.}
\label{triangulo}
\end{figure}

where $D(K)$ is given by Eq.~(\ref{bosonicop}) and S(K) is the fermion propagator 
\begin{equation}
    \label{ec:propfermion}
    S(\mathbf{k})=i\frac{\slashed{k}+m_{f}}{k^{2}-m_{f}^{2}},
\end{equation}
therefore, considering all diagrams, the correction of the self-coupling $g$ to one-loop order is given by
\begin{equation}
   g_{\mbox{\small{eff}}}=g
   \left[1 + 4 \frac{G(P_i;m^2)}{-ig}
   \right].
\label{ggeff}
\end{equation}
In order to proceed, we have to compute $G(P_i;m^2)$. Using trace algebra and applying the Dirac equation in Eq.(\ref{ec:feynman1}), we get
\begin{eqnarray}
    \label{ec:feynman5}
   && G(P_i,m^2)=(-ig)(-ig^{2})\int\frac{d^{4}k}{(2\pi)^{4}} \nonumber \\
    &\times&\bigg[\frac{1}{[(P_{1}-K)^{2}-m_{f}^{2}][(P_{2}-K)^{2}-m_{f}^{2}]}\nonumber \\
    &+&\frac{m_{\pi}^{2}}{[(P_{1}-K)^{2}-m_{f}^{2}][(P_{2}-K)^{2}-m_{f}^{2}][K^{2}-m_{\pi}^{2}]}\bigg], \nonumber \\
\end{eqnarray}
In order to apply the imaginary-time formalism of finite temperature field theory, we make the substitutions $p_{0}\rightarrow i\tilde{\omega}_{m}+\mu$ for fermion part, $k_{0}\rightarrow i\omega_{n}$  for boson part, where $\omega_n=2n\pi T$, and 
\begin{equation}
\label{ec:matsubara}
    \int \frac{d^{4}k}{(2\pi)^{4}}\rightarrow i T\sum_{n}\int \frac{d^{3}k}{(2\pi)^{3}},
\end{equation}
therefore, we obtain
\begin{eqnarray}
  &&  G(P_i,m^2)=(-ig)(g^{2}) T\sum_{n}\int \frac{d^{3}k}{(2\pi)^{3}}\nonumber \\
    &\times&\bigg[\frac{1}{[(\tilde{\omega}_{1}-i\mu+\omega_{n})^{2}+E_{1}^{2}][(\tilde{\omega}_{2}-i\mu+\omega_{n})^{2}+E_{2}^{2}]}\nonumber\\
  &&  -\bigg(\frac{m_{\pi}^{2}}{[(\tilde{\omega}_{1}-i\mu+\omega_{n})^{2}+E_{1}^{2}][(\tilde{\omega}_{2}-i\mu+\omega_{n})^{2}+E_{2}^{2}]}\nonumber\\
  && \times \frac{1}{[\omega_{n}^{2}+E^{2}]}\bigg)\bigg],\nonumber \\
    \label{ec:feynman6}
\end{eqnarray}
where 
\begin{eqnarray}
E_1^2&=&(\textbf{p}_1-\textbf{k})^2+m_f^2 \nonumber \\
E_2^2&=&(\textbf{p}_2-\textbf{k})^2+m_f^2 \nonumber \\
E^2&=&\textbf{k}^2+m_{\pi}^2.
\end{eqnarray}
Both sums in Eq.~(\ref{ec:feynman6}) are calculated using Ref.~\cite{LeBellac}. We obtain 
\begin{eqnarray}
\text{I}&=&\sum_n\frac{1}{[(\tilde{\omega}_{1}-i\mu+\omega_{n})^{2}+E_{1}^{2}][(\tilde{\omega}_{2}-i\mu+\omega_{n})^{2}+E_{2}^{2}]} \nonumber \\
&=&\sum_{s1,s2}\frac{s_1 s_2}{4E_1 E_2} \frac{(-1+\tilde{f}(s_2 E_2+\mu)+\tilde{f}(s_1 E_1-\mu))}{ i(\tilde{\omega}_1-\tilde{\omega}_2) - s_1 E_1 - s_2 E_2}, \nonumber \\ \label{summ1} 
\end{eqnarray}
\begin{eqnarray}
\text{II}&=&\sum_n\frac{m_{\pi}^{2}}{[(\tilde{\omega}_{1}-i\mu+\omega_{n})^{2}+E_{1}^{2}]} \nonumber \\
&\times&\frac{m_{\pi}^{2}}{[(\tilde{\omega}_{2}-i\mu+\omega_{n})^{2}+E_{2}^{2}][\omega_{n}^{2}+E^{2}]} \nonumber \\
&=&\sum_{s,s_{1},s_{2}}\frac{-ss_{1}s_{2}}{8EE_{1}E_{2}}\frac{1}{[i(\tilde{\omega}_{1}-\tilde{\omega}_{2})+s_{2}E_{2}-s_{1}E_{1}]}\nonumber \\
&\times&\bigg[\frac{1+f(sE)-\tilde{f}(s_{1}E_{1}-\mu)}{i\tilde{\omega}_{1}-\mu-sE-s_{1}E_{1}}\nonumber\\
&-&\frac{1+f(sE)-\tilde{f}(s_{2}E_{2}-\mu)}{i\tilde{\omega}_{2}-\mu-sE-s_{2}E_{2}}\bigg],\label{summ2}
\end{eqnarray}

\begin{figure}[t]
\begin{center}
\includegraphics[scale=0.58]{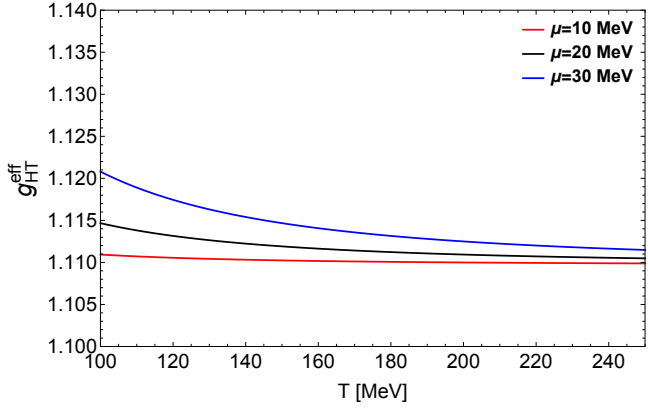}
\end{center}
\caption{Effective fermion-boson coupling $g_{\text{eff}}$, in the high-temperature approximation, as a function of $T$.}
\label{fig7}
\end{figure}

To study the high (HT)- and low-temperature (LT) regimes, we analyze Eq.~(\ref{summ1}) and Eq.~(\ref{summ2}) in each regime. First, we compute the high temperature regime. For the case of $\text{I}_{\text{HT}}$, the dominant terms behave as $\sim T^{2}$. This behavior is obtained for the case when $s_{1}=-s_{2}$ in (\ref{summ1}). Considering only the matter piece, we get
\begin{eqnarray}
    \text{I}_{\text{HT}}&=&-\frac{1}{4E_{1}E_{2}}\bigg[\frac{1-\tilde{f}(E_{2}-\mu)+\tilde{f}(E_{1}-\mu)}{i(\tilde{\omega}_{1}-\tilde{\omega}_{2})-E_{1}+E_{2}}\nonumber \\ 
    &+&\frac{\tilde{f}(E_{2}+\mu)+1-\tilde{f}(E_{1}+\mu)}{i(\tilde{\omega}_{1}-\tilde{\omega}_{2})+E_{1}-E_{2}}\bigg],
\end{eqnarray}
where have we used $1+f(-E)=-f(E)$ and $\tilde{f}(-E_{1}-\mu)=1-\tilde{f}(E_{1}+\mu)$. 
Taking $E_{1}=E_{2}$, we obtain
\begin{gather}
    \text{I}_{\text{HT}}=\frac{1}{4E_{1}E_{2}i(\tilde{\omega}_{1}-\tilde{\omega}_{2})}[1-1]=0 \label{cadc}.
\end{gather}
Therefore, $\text{I}_{\text{HT}}$ does not contribute to terms of order $T^{2}$. Now we compute $\text{II}_{\text{HT}}$. Once again, terms that behave like $\sim T^{2}$ are obtained when $s=-s_{1}=-s_{2}$ in Eq,~(\ref{summ2}). Therefore we get,
\begin{eqnarray}
    \text{II}_{\text{HT}}&=&\frac{-1}{8EE_{1}E_{2}}\bigg(\frac{1}{i\tilde{\omega}_{1}-\tilde{\omega}_{2})+E_{1}-E_{2}}\nonumber \\
    &\times&\bigg[\frac{f(E)+\tilde{f}(E_{1}+\mu)}{p_{1}^{0}-E+E_{1}}-\frac{f(E)+\tilde{f}(E_{2}+\mu)}{p_{2}^{0}-E+E_{2}}\bigg]\nonumber\\
    &+&\frac{1}{i(\tilde{\omega}_{1}-\tilde{\omega}_{2})-E_{1}+E_{2}}\nonumber \\
    &\times&\bigg[\frac{f(E)+\tilde{f}(E_{1}-\mu)}{p_{1}^{0}+E-E_{1}}-\frac{f(E)+\tilde{f}(E_{2}-\mu)}{p_{2}^{0}+E-E_{2}}\bigg]\bigg),
    \nonumber\\[1ex]
    \label{ec:3propHTL1}
\end{eqnarray}
where have we used $1+f(-E)=-f(E)$ and $\tilde{f}(-E_{1}-\mu)=1-\tilde{f}(E_{1}+\mu)$,  $p_{1}^{0}=i\tilde{\omega}_1-\mu$ and $p_{2}^{0}=i\tilde{\omega}_2-\mu$. 
Taking $E_{1}=E_{2}$, we obtain
\begin{eqnarray}
    \text{II}_{\text{HT}}
    &=&\frac{1}{8EE_{1}^{2}}\bigg(\frac{f(E)+\tilde{f}(E_{1}+\mu)}{(p_{1}^{0}-E+E_{1})(p_{2}^{0}-E+E_{1})}\nonumber \\ &+&\frac{f(E)+\tilde{f}(E_{1}-\mu)}{(p_{1}^{0}+E-E_{1})(p_{2}^{0}+E_{1}-E)}\bigg).
    \label{ec:3propHTL2}
\end{eqnarray}
Taking $p_{1}^{0}=0$ and   $p_{2}^{0}=0$, we get
\begin{gather}
    \text{II}_{\text{HT}}=\frac{k^{2}}{8EE_{1}^{2}}\bigg(\frac{2f(E)+\tilde{f}(E_{1}+\mu)+\tilde{f}(E_{1}-\mu)}{m_{\pi}^{4}}\bigg).
    \label{ec:3propHTL 4}
\end{gather}
\begin{figure}[t]
\begin{center}
\includegraphics[scale=0.58]{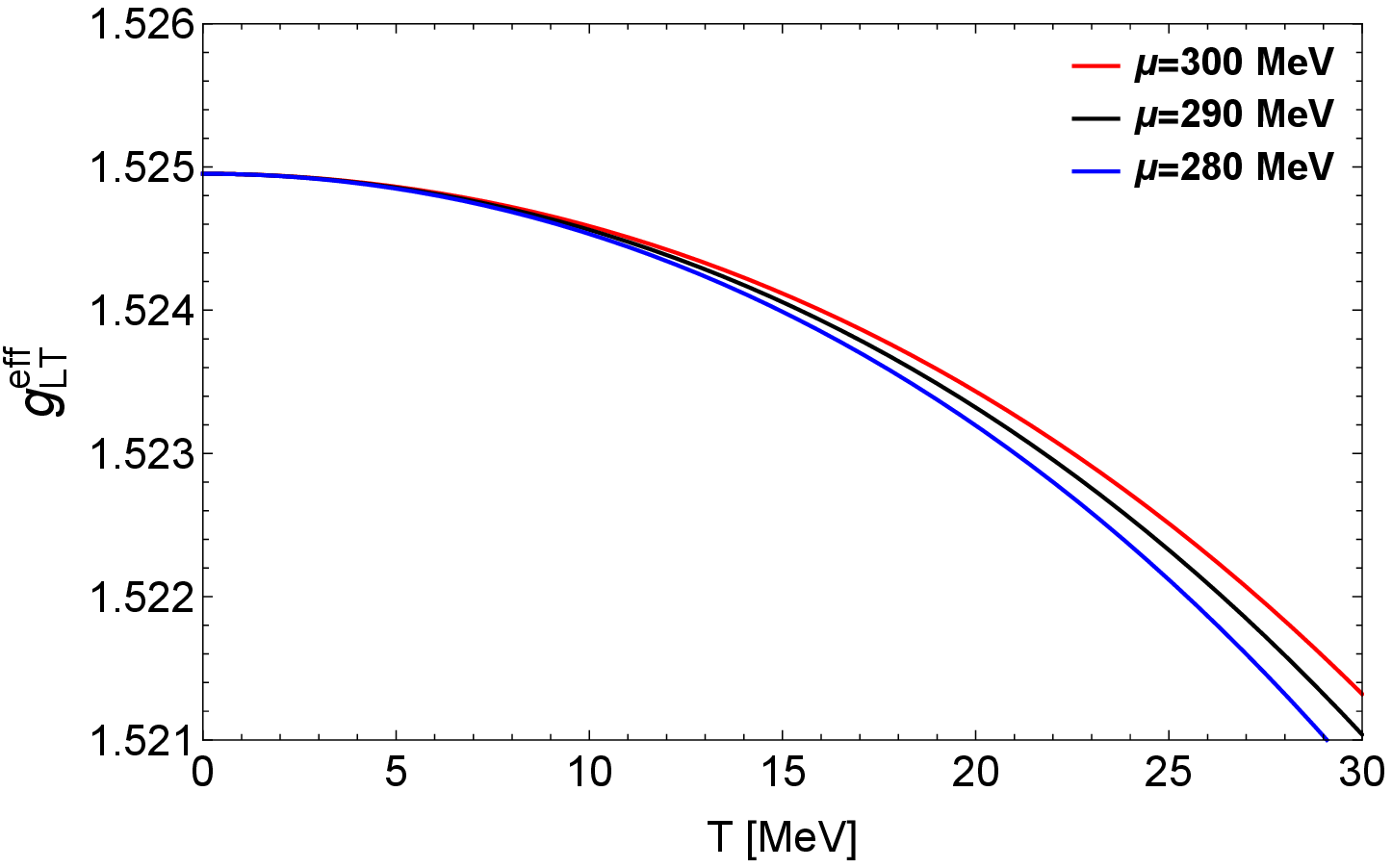}
\end{center}
\caption{Effective fermion-boson coupling $g_{\text{eff}}$, in the low-temperature approximation, as a function of $T$.}
\label{fig8}
\end{figure}

Therefore, using Eq.~(\ref{ec:3propHTL 4}) and Eq.~(\ref{cadc})  in Eq.
~(\ref{ec:feynman6}), we obtain
\begin{eqnarray}
  &&G(0,m^2)_{\text{HT}}=  (-ig)(g^{2})\int\frac{d^{3}k}{(2\pi)^{3}}\frac{-k^{2}}{8EE_{1}^{2}} \nonumber \\
  &\times&\bigg(\frac{2f(E)+\tilde{f}(E_{1}+\mu)+\tilde{f}(E_{1}-\mu)}{m_{\pi}^{2}}\bigg)\nonumber\\
    &=&-(-ig)(g^{2})\frac{4}{16\pi^{2}m_{\pi}^{2}}\int_{0}^{\infty}\frac{dk\,k^{4}}{\sqrt{k^{2}+m_{\pi}^{2}}k^{2}}\nonumber \\
    &\times&(2f(\sqrt{k^{2}+m_{\pi}^{2}})+\tilde{f}(k+\mu)+\tilde{f}(k-\mu)\nonumber\\
    &=&\frac{(-ig)(g^{2})}{4\pi^{2}m_{\pi}^{2}}\bigg[2\int_{0}^{\infty}\frac{dk\,k^{2}}{\sqrt{k^{2}+m_{\pi}^{2}}}f(\sqrt{k^{2}+m_{\pi}^{2}})\nonumber \\
    &+&\int_{0}^{\infty}dk\frac{k^{2}}{\sqrt{k^{2}+m_{\pi}^{2}}}(\tilde{f}(k+\mu)+\tilde{f}(k-\mu))\bigg].
    \label{ec:3propHTL5}
\end{eqnarray}
The first integral in Eq.~(\ref{ec:3propHTL5}) is computed using Eq.~(\ref{handf}) and Eq.~(\ref{diffeq}), we get
\begin{eqnarray}
  &&G(0,m^2)_{\text{HT}} =\frac{(-ig)(g^{2})}{4\pi^{2}m_{\pi}^{2}}\bigg[-m_{\pi}\pi T \nonumber \\
  &+&\int_{0}^{\infty}dk\frac{k^{2}}{\sqrt{k^{2}+m_{\pi}^{2}}}(\tilde{f}(k+\mu)+\tilde{f}(k-\mu))\bigg].
    \label{ec:3propHTL6}
\end{eqnarray}
In the high-temperature case, we can approximate  $\sqrt{k^{2}+m_{\pi}^{2}}\sim k$ in the denominator of the previous integral. Then, we have an analytical integral of the kind
\begin{equation}
    \int_{0}^{\infty}\frac{k}{e^{\frac{k\pm\mu}{T}}+1}=-T^2 Li_{2}(-e^{\mp \mu/T}).
    \label{ec:3propHTL7}
\end{equation}
Hence, we obtain
\begin{eqnarray}
  G(0,m^2)_{\text{HT}}&=&\frac{(-ig)(g^{2})}{4\pi^{2}m_{\pi}^{2}}T[-m_{\pi}\pi-T(Li_{2}(-e^{-\frac{-\mu}{T}})\nonumber \\
  &-&Li_{2}(-e^{-\frac{\mu}{T}}))].\label{ghtt}
\end{eqnarray}
Finally, using Eq.~(\ref{ghtt}) into Eq.~(\ref{ggeff}), we get 
\begin{align}
    g^{\text{eff}}_{HT}=g&\Bigg[1-4g^2\frac{T^2}{4\pi^{2} \Pi_b^{HT}}\Big[\frac{\pi(\Pi_b^{HT})^{1/2} }{T}\nonumber \\ &+Li_{2}(-e^{\frac{-\mu_{q}}{T}})+Li_{2}(-e^{\frac{\mu_{q}}{T}})\Big]\Bigg].
    \label{geffHT1}
\end{align}

Now we proceed to compute the low temperature regime. When $T\to 0$, the Fermi-Dirac distribution behaves as
\begin{eqnarray}
    \label{ec:aproximacionFM-1}
    \tilde{f}(x+\mu)&=&\frac{1}{e^{\frac{x+\mu}{T}}+1}\rightarrow0 \nonumber \\ \tilde{f}(x-\mu)&=&\frac{1}{e^{\frac{x-\mu}{T}}+1}\rightarrow\Theta(\mu-x).
\end{eqnarray}
Using Eq.~(\ref{ec:aproximacionFM-1}) into Eq.~(\ref{summ1}), we obtain
\begin{widetext}

\begin{gather}
    \text{I}_{\text{LT}}
    =\frac{1}{4E_{1}E_{2}}\bigg[\frac{\Theta(\mu-E_{1})}{i(\tilde{\omega}_{1}-\tilde{\omega}_{2})-E_{1}-E_{2}}+\frac{-1}{i(\tilde{\omega}_{1}-\tilde{\omega}_{2})-E_{1}-E_{2}}+\frac{\Theta(\mu-E_{2})}{i(\tilde{\omega}_{1}-\tilde{\omega}_{2})-E_{1}+E_{2}}+\frac{-\Theta(\mu-E_{1})}{i(\tilde{\omega}_{1}-\tilde{\omega}_{2})-E_{1}+E_{2}}+\nonumber\\[1ex]
    \frac{1}{i(\tilde{\omega}_{1}-\tilde{\omega}_{2})+E_{1}+E_{2}}-\frac{\Theta(\mu-E_{2})}{i(\tilde{\omega}_{1}-\tilde{\omega}_{2})+E_{1}+E_{2}}\bigg].
    \label{ec:LTL2prop}
\end{gather}

\end{widetext}
Setting $E_1=E_2$, we get
\begin{equation}
\text{I}_{\text{LT}}=\frac{1}{4E_{1}^{2}}\bigg[\frac{\Theta(\mu-E_{1})-1}{i(\tilde{\omega}_{1}-\tilde{\omega}_{2})-2E_{1}}+\frac{1-\Theta(\mu-E_{1})}{i(\tilde{\omega}_{1}-\tilde{\omega}_{2})+2E_{1}}\bigg].
    \label{ec:LTL2prop2}
\end{equation}
Taking the limit $\tilde{\omega}_{1}=\tilde{\omega}_{2}=0$, we obtain
\begin{equation}
\text{I}_{\text{LT}}=\frac{1}{4E_{1}^{3}}[1-\Theta(\mu-E_{1})].
    \label{ec:LTL2prop3}
\end{equation}
Now, we analyze II. For this case we have to consider that for $T\rightarrow0$, the Bose-Einstein distribution behaves as   
\begin{eqnarray}
    \label{ec:Bose-Einsteinlimite}
    f(E)&=&\frac{1}{e^{E/T}-1}\rightarrow0  \nonumber \\ 1+f(-E)&=&\frac{1}{1-e^{E/T}}\rightarrow0.
\end{eqnarray}
Therefore, using Eq.~(\ref{ec:aproximacionFM-1}) and Eq.~(\ref{ec:Bose-Einsteinlimite}), and proceeding in the same manner as in the previous calculations, {\it i.e.} setting $E_1=E_2$ and $p_{1}^{0}=p_{2}^{0}=\tilde{\omega}_{2}=\tilde{\omega}_{1}=0$, we obtain
\begin{gather}
\text{II}_{\text{LT}}=\frac{2E_{1}+E}{4E_{1}^{3}E(E_{1}+E)^{2}}+\frac{(3E_{1}^{3}E-E^{3})\tilde{f}(E_{1}-\mu)}{4E_{1}^{3}E(E_{1}^{2}-E^{2})^{2}}
\label{ec:LTL3prop2}.
\end{gather}
Hence, considering only the matter part of Eq.~(\ref{ec:LTL2prop3}) and  Eq.~(\ref{ec:LTL3prop2}) in Eq.~(\ref{ec:feynman6}), we get
\begin{eqnarray}
  &&  G(0,m^2)_{\text{LT}}=(-ig)(g^{2}) \int \frac{d^{3}k}{(2\pi)^{3}} \bigg[ \frac{-1}{4E_{1}^{3}}\Theta(\mu-E_{1}) \nonumber \\
    &+&\frac{(3E_{1}^{3}E-E^{3})\tilde{f}(E_{1}-\mu)}{4E_{1}^{3}E(E_{1}^{2}-E^{2})^{2}}\bigg].  
    \label{ltlttt}
\end{eqnarray}
We have two integrals to calculate
\begin{equation}
 G_{\text{I}}= \int \frac{d^{3}k}{(2\pi)^{3}}  \frac{-1}{4E_{1}^{3}}\Theta(\mu-E_{1}) ,
\end{equation}
and
\begin{equation}
  G_{\text{II}}=\int \frac{d^{3}k}{(2\pi)^{3}} 
  \frac{(3E_{1}^{3}E-E^{3})\tilde{f}(E_{1}-\mu)}{4E_{1}^{3}E(E_{1}^{2}-E^{2})^{2}}.
\end{equation}
The integral $G_{\text{I}}$ is straightforward,
\begin{eqnarray}
    G_{\text{I}}&=&-\int\frac{d^{3}k}{(2\pi)^{3}}\frac{\Theta(\mu-E_{1})}{E_{1}^{3}}\nonumber \\
    &=&-\frac{1}{8\pi^{2}}\int^{\infty}_{0}\frac{k^{2}\Theta(\mu-E_{1})}{(k^{2}+m_{f}^{2})^{3/2}}dk, \nonumber \\
    &=&-\frac{1}{8\pi^{2}}\int^{\sqrt{\mu^{2}-m_{f}^{2}}}_{0}\frac{k^{2}\Theta(\mu-E_{1})}{(k^{2}+m_{f}^{2})^{3/2}}dk \nonumber \\
    &=&-\frac{1}{8\pi^{2}}\bigg[\log\bigg(\frac{\mu+\sqrt{\mu^{2}-m_{f}^{2}}}{m_{f}}\bigg)-\frac{\sqrt{\mu^{2}-m_{f}^{2}}}{\mu}\bigg]. \nonumber \\
    \label{ec:LTL2propE}
\end{eqnarray}
For $G_{\text{II}}$, we use Eq.~(\ref{ec:partitionfunction0}) and obtain
\begin{eqnarray}     G_{\text{II}}&=&\frac{m_{\pi}^{2}}{2\pi^{2}}\int_{0}^{\sqrt{\mu^{2}-m_{f}^{2}}}dk k^{2}\frac{2k^{2}+3m_{f}^{2}-m_{\pi}^{2}}{4(k^{2}+m_{f}^{2})^{3/2}(m_{f}^{2}-m_{\pi}^{2})^{2}}\nonumber\\[1ex]
    &=&\frac{m_{\pi}^{2}}{16\pi^{2}\mu(m_{f}^{2}-m_{\pi}^{2})^{2}}\nonumber \\
    &\times&\bigg[-2\mu m_{\pi}^{2}\ln\bigg(\frac{\mu+\sqrt{\mu^{2}-m_{f}^{2}}}{m_{f}}\bigg)\nonumber \\
    &+&2\sqrt{\mu^{2}-m_{f}^{2}}(\mu^{2}-m_{f}^{2}+m_{\pi}^{2})\bigg].
    \label{ec:LTL3propfinal}
\end{eqnarray}
Therefore, using Eq.~(\ref{ec:LTL2propE}) and Eq.~(\ref{ec:LTL3propfinal}) into Eq.~(\ref{ltlttt}), we get
\begin{eqnarray}
  && G(0,m^2,T\rightarrow0)_{\text{LT}}=(-ig)(g^{2})\bigg[\frac{\sqrt{\mu^{2}-m_{f}^{2}}}{8\pi^{2}\mu}\nonumber \\
   &-&\frac{1}{8\pi^{2}}\bigg[\log\bigg(\frac{\mu+\sqrt{\mu^{2}-m_{f}^{2}}}{m_{f}}\bigg)\bigg]-\nonumber\\
 &&   \frac{m_{\pi}^{2}}{16\pi^{2}\mu(m_{f}^{2}-m_{\pi}^{2})^{2}}\nonumber \\
 &\times&\bigg[-2\mu m_{\pi}^{2}\ln\bigg(\frac{\mu+\sqrt{\mu^{2}-m_{f}^{2}}}{m_{f}}\bigg)\nonumber \\
 &+&2\sqrt{\mu^{2}-m_{f}^{2}}(\mu^{2}-m_{f}^{2}+m_{\pi}^{2})\bigg]\bigg]\equiv G_{0}(\mu).
    \label{ec:G}
\end{eqnarray}
Notice that Eq.~(\ref{ec:G}) is an approximation valid for $T\sim0$. To find the next order terms in $T$, we use a series around $T\sim 0$ \cite{chilenos}, 
\begin{equation}
    G(0,m^2)_{\text{LT}}=G_{0}(\mu)+\frac{\pi^{2}T^{2}}{6}\frac{\partial^{2}G_{0}(\mu+xT)}{\partial(xT)^2}\biggr\rvert_{T=0}+..., 
\end{equation}
\begin{widetext}
Hence we obtain
\begin{eqnarray}
 &&G(0,m^2)_{\text{LT}}=(-ig)(g^{2})\bigg[\frac{\sqrt{\mu^{2}-m_{f}^{2}}}{8\pi^{2}\mu}-\frac{1}{8\pi^{2}}\bigg[\ln\bigg(\frac{\mu+\sqrt{\mu^{2}-m_{f}^{2}}}{m_{f}}\bigg)\bigg]-  \frac{m_{\pi}^{2}}{16\pi^{2}\mu(m_{f}^{2}-m_{\pi}^{2})^{2}}\nonumber \\
  &\times&\bigg[-2\mu m_{\pi}^{2}\ln\bigg(\frac{\mu+\sqrt{\mu^{2}-m_{f}^{2}}}{m_{f}}\bigg)+2\sqrt{\mu^{2}-m_{f}^{2}}(\mu^{2}-m_{f}^{2}+m_{\pi}^{2})\bigg]\nonumber \\
  &+&\frac{T^{2}}{48}\frac{(-2m_{f}^{6}+m_{f}^{4}(\mu^{2}+2m_{\pi}^{2})-\mu^{2}m_{f}^{2}m_{\pi}^{2}-2\mu^{4}m_{\pi}^{2})}{\mu^{3}\sqrt{\mu^{2}-m_{f}^{2}}(m_{f}^{2}-m_{\pi}^{2})^{2}}\bigg]. \nonumber \\
    \label{ec:LTLfinal}
\end{eqnarray}

Finally, using Eq.~(\ref{ec:LTLfinal}) into Eq.~(\ref{ggeff}), we get 
\begin{align}
    &g^{\text{eff}}_{LT}=g\Bigg[1+4g^2\bigg[\frac{\sqrt{\mu_q^{2}-m_f^2}}{8\pi^{2}\mu_q}-\frac{1}{8\pi^{2}}\ln\bigg(\frac{\mu_q+\sqrt{\mu_q^{2}-m_f^2}}{m_f}\bigg)+\frac{2(\Pi_b^{LT})^2}{16\pi^{2}(m_{f}^{2}-\Pi_b^{LT})^{2}}\bigg[ \ln\bigg(\frac{\mu_q+\sqrt{\mu_q^{2}-m_{f}^{2}}}{m_{f}}\bigg)\nonumber \\
    &-\bigg(1-\frac{m_f^2}{\mu_q^{2}}\bigg)^{1/2}(1+\frac{\mu_q^{2}-m_{f}^{2}}{\Pi_b^{LT}})\bigg]+\frac{T^{2}}{48}\frac{(-2m_{f}^{6}+m_{f}^{4}(\mu_q^2+2\Pi_b^{LT})-\mu_q^2m_f^{2}\Pi_b^{LT}-2\mu_q^4\Pi_b^{LT})}{\mu_q^3\sqrt{\mu_q^2-m_{f}^{2}}(m_{f}^{2}-\Pi_b^{LT})^{2}}\bigg] \Bigg].
    \label{geffLTtt1}
\end{align}
\end{widetext}

\

\end{document}